\documentclass{aa}
\usepackage{graphicx}
\usepackage{psfig}
\usepackage{latexsym}
\usepackage{amssymb}
\usepackage{lscape}
\usepackage{amsmath}

\newcommand{\MC}{\multicolumn}
\newcommand{\kms}{km~s$^\mathrm{-1}$}
\newcommand{\sunn}{$_{\odot}$}

\newcounter{qub}
\begin{document}

\title{Possibly interacting Vorontsov-Velyaminov galaxies}
\subtitle{II. The 6\,m telescope spectroscopy of VV~080, 131, 499, 523 and 531}

\author{Pustilnik S.\inst{1,4} \and
Zasov A.\inst{2} \and
Kniazev A.\inst{1,3,4} \and
Pramskij A.\inst{1,4} \and
Ugryumov A.\inst{1,4} \and
Burenkov A.\inst{1,4}
}

\offprints{S. Pustilnik, \email{sap@sao.ru}}

\institute{
Special Astrophysical Observatory, Nizhnij Arkhyz,
Karachai-Circassia, 369167  Russia
\and
Sternberg Astronomical Institute of the Moscow State
University, Universitetsky Pr., 13, Moscow, 119899, Russia
\and Max Planck Institut f\"{u}r Astronomie, K\"{o}nigstuhl 17, D-69117,
Heidelberg, Germany
\and
Isaac Newton Institute of Chile, SAO Branch
 }

\date{Received 21 August 2002 / Accepted 29 November 2002}

\abstract{In continuation of the program formulated in Paper I
we present results of long-slit spectroscopy with the Russian 6\,m
telescope of five more objects from the Vorontsov-Velyaminov
(hereafter VV) Atlas and Catalogue of Interacting galaxies. These
are the galaxies for which the interaction is not evident,
although all of them are classified as multiple systems (usually
as ``nests'', ``chains'' and similar systems). The
spectrophotometry data enable us to derive for all galaxies
abundances of O, and for some of them abundances of N, Ne and S.
For two of them chemical abundances are given for the first time.
In spectra of three of the studied galaxies [\ion{O}{iii}]
$\lambda$4363 line is measured, and T$_\mathrm{e}$ and oxygen
abundance are derived by the classical method. For the two others,
empirical methods are used. For all 5 galaxies, the radial velocity
distribution along the slit was obtained in the H$\alpha$-line. The
studied galaxies represent a rather mixed sample: from very low
luminosity irregular galaxies, like VV~499 (DDO~053), to rather
bright VV~523 (NGC~3991). Their metallicities vary from $Z
\sim$1/25~$Z$\sunn\ for VV~499 to $\sim$1/2~$Z$\sunn\ for VV~523.
The morphology of these galaxies ranges between typical dIrr
(VV~080)  through ring-like (VV~131) to clumpy Irr (VV~523).
Position--Velocity diagrams in the H$\alpha$-line along the galaxy
body imply the existence of large-scale ionized gas
outflows/supershells around the sites of intense current and/or
recent SF activity. Sizes of supershells vary in the range of
several hundred pc to $\sim$2 kpc. For all studied galaxies we
examine their local environment and indicate the nearest
neighbouring objects capable of inducing the observed enhanced star
formation.
}

\authorrunning{S.A. Pustilnik et al.}

\titlerunning{Study VV galaxies. II.}

\maketitle


\begin{table*}
\begin{center}
\caption{\label{Tab2} Journal of observations}
\begin{tabular}{llrrrccc} \\ \hline
\MC{1}{c}{ Galaxy } &
\MC{1}{c}{ Date } &
\MC{1}{c}{ Grating } &
\MC{1}{c}{ Exposure } &
\MC{1}{c}{ Wavelength } &
\MC{1}{c}{ Dispersion } &
\MC{1}{c}{ PA } \\

\MC{1}{c}{ Name } &  &
\MC{1}{c}{ [grooves/mm] } &
\MC{1}{c}{ time [s] } &
\MC{1}{c}{ Range [\AA] } &
\MC{1}{c}{ [\AA/pixel] } &
\MC{1}{c}{ [Degree] } \\

\MC{1}{c}{ (1) } &
\MC{1}{c}{ (2) } &
\MC{1}{c}{ (3) } &
\MC{1}{c}{ (4) } &
\MC{1}{c}{ (5) } &
\MC{1}{c}{ (6) } &
\MC{1}{c}{ (7) } \\
\hline
\\[-0.3cm]
VV~080 &  04.09.1999  & 1302 & 900            & 6000--7200 & 1.2 & ~~45  \\
VV~080 &  01.11.1999  & 325  & 1200           & 3700--8000 & 4.6 & ~~55 \\
VV~080 &  01.11.1999  & 325  & 1200           & 3700--8000 & 4.6 & 125 \\   \hline
VV~531 &  04.09.1999  & 1302 &  1200          & 6000--7200 & 1.2 & ~~42 \\
VV~531 &  04.09.1999  & 1302 &  900           & 6000--7200 & 1.2 & ~~96 \\
VV~531 &  02.11.1999  & 651  &  900           & 3700--6000 & 2.4 & ~~42 \\
VV~531 &  02.11.1999  & 651  &  1200          & 3700--6000 & 2.4 & ~~96 \\  \hline
VV~499 &  12.02.1999  & 325  &  600           & 3700--8000 & 4.6 & 145  \\
VV~499 &  12.02.1999  & 325  & 1200           & 3700--8000 & 4.6 & 145  \\
VV~499 &  25.04.1999  & 1302 &  2$\times$1200 & 6000--7200 & 1.2 & ~~98 \\
VV~499 &  25.04.1999  & 1302 & 900            & 6000--7200 & 1.2 & 145  \\
VV~499 &  06.10.1999  & 651  &  2$\times$1200 & 3700--6000 & 2.4 & 145  \\
VV~499 &  05.10.2000  & 651  &  2$\times$1800 & 3700--6000 & 2.4 & 140  \\
VV~499 &  26.12.2001  & 1302 &  2$\times$1800 & 6000--7200 & 1.2 & 100,145  \\  \hline
VV~131 &  13.02.1999  & 325  &  600           & 3700--8000 & 4.6 & 158  \\
VV~131 &  03.11.1999  & 1302 &  1800          & 6000--7200 & 1.2 & 158  \\
VV~131 &  26.12.2001  & 1302 &  1800,900      & 6000--7200 & 1.2 & 46,103 \\  \hline
VV~523 &  12.07.1999  & 325  &  2$\times$900  & 3700--8000 & 4.6 & ~~33 \\
VV~523 &  14.07.1999  & 1302 &  1200          & 6000--7200 & 1.2 & ~~33 \\
\hline \\[--0.2cm]

\end{tabular}
\end{center}
\end{table*}

\begin{table*}[hbtp]
\centering{
\caption{\label{Tab1} Main parameters of VV galaxies studied}

\begin{tabular}{lrrrrrr} \hline\hline
\rule{0pt}{10pt}
Parameter                     & VV~080               & VV~531               & VV~499               & VV~131                & VV~523                \\ \hline
Other name                    & NGC~14               & NGC~1156             & DDO~053              & UGC~4874              & NGC~3991              \\
$\alpha_{2000}$               & 00$^h$08$^m$46.2$^s$ & 02$^h$59$^m$42.5$^s$ & 08$^h$34$^m$06.9$^s$ & 09$^h$15$^m$17.43$^s$ & 11$^h$57$^m$31.5$^s$  \\
$\delta_{2000}$               & +15\degr49\arcmin01\arcsec & +25\degr14\arcmin15\arcsec  & +66\degr10\arcmin55\arcsec  &+48\degr39\arcmin55\arcsec & +32\degr20\arcmin25\arcsec \\
$B_{\rm tot}^{(1)}$            & 12.71$\pm$0.16       & 12.32$\pm$0.13       & 14.48$\pm$0.17       & 16.0$\pm$0.5          & 13.50$\pm$0.20        \\
$A_{\rm B}^{(1)}$ (mag)        & 0.33                 & 0.97                 &  0.16                & 0.07                  & 0.10                  \\
$(B-V)_{\rm tot}^{(1)}$        & 0.58$\pm$0.23        & 0.58$\pm$0.18        & 0.35$\pm$0.24        & ---                   & 0.39$\pm$0.28         \\
$V_{\rm Hel}^{(1)}$ (\kms)     & 865$\pm$1            & 375$\pm$1            & --19$\pm$10          & 2786$\pm$5            & 3192$\pm$5            \\
Distance (Mpc)$^{(2)}$          & 11.7                 & 7.8$^{(3)}$            & 3.56$^{(4)}$           & 39.6                  & 45.3                  \\
$M_{\rm B}^{0}$$^{(2)}$      & --18.0               & --17.7               & --13.6               & --17.1                & --20.9                \\
D$_{\rm 25}$  (\arcsec)  & 151$\pm$8            & 190$\pm$7            &  67$\pm$9            & 46$\pm$8              &  86$^{(8)}$             \\
D$_{\rm 25}$  (kpc)$^{(2)}$  & 9.3                  & 7.2                  &  1.3                 & 8.3                   &  18.9                 \\
Axial ratio (b/a)$^{(1)}$         & 0.75                 & 0.76                 &  0.87                & 0.82                  & 0.33                  \\
12+$\log$(O/H)  \              & 8.6$^{9}$            & 8.17$\pm$0.08        & 7.52$\pm$0.08        & 7.85$\pm$0.06         & 8.65$^{(10)}$             \\
HI integr.flux$^{(1)}$ (Jy$\cdot$\kms)   & 12.18                & 72.7$^5$             & 23.7                 & 9.33                   & 17.0                  \\
$W_{\rm 20}$~\kms~$^{(11)}$& 114$\pm$17           & 103$\pm$3$^{(6)}$       & 30$^{(7)}$             & 114$\pm$7             & 245$\pm$12            \\
M(HI) (10$^{8}~M$\sunn)       & 3.9                  & 10.4                 & 0.71                 & 34.5                  & 82.3                  \\
M(HI)/$L_{\rm B}$$^{(2)}$           & 0.16                 & 0.56                 & 1.7                  & 3.2                   & 0.23                   \\
\hline\hline
\multicolumn{6}{l}{$B_{\rm tot}$ -- total blue magnitude; $M_{\rm B}^{0}$ -- extinction corrected absolute magnitude} \\
\multicolumn{6}{l}{D$_{\rm 25}$ -- Diameter at the surface brightness
$\mu_{\rm B}$=25$^{m}$/$\Box$\arcsec\ from LEDA; $A_{\rm B}$ -- Galactic extinction in blue} \\
\multicolumn{6}{l}{$^{(1)}$\,\ Data from NED; $B$-mag. from UGC and VV catalogs} \\
\multicolumn{6}{l}{$^{(2)}$\,\ See explanations in  section~\ref{Individual_prop}; $^{(3)}$\,\ from Karachentsev et al. (\cite{Kara96}); $^{(4)}$\, Karachentsev et al. (\cite{Kara02})} \\
\multicolumn{6}{l}{$^{(5)}$\,\ Haynes et al. (\cite{Haynes98}); $^{(6)}$\,\ $W_{\rm 20}$ from Broeils \& van Woerden (\cite{BvW94})} \\
\multicolumn{6}{l}{$^{(7)}$\,\ this is $W_{\rm 50}$ from Huchtmeier \& Richter (\cite{Huchtmeier89}); $^{(8)}$\,\ from Hecquet et al. \cite{Hecquet95} } \\
\multicolumn{6}{l}{$^{(9)}$\,\ Determined by combination of empirical methods (see section~\ref{VV080abun})} \\
\multicolumn{6}{l}{$^{(10)}$\,\ Determined with the use of the empirical method by Pilyugin (\cite{Pilyugin01})} \\
\multicolumn{6}{l}{$^{(11)}$\,\ $W_{\rm 20}$ from Huchtmeier \& Richter (\cite{Huchtmeier89})} \\
\end{tabular}
}
\end{table*}

\section{Introduction}

As already noticed in the introduction to Paper I (Zasov et al.
\cite{Zasov00}), among definitely interacting galaxies in close
pairs or groups, which were (in most cases) discovered by
Vorontsov-Velyaminov and included in his Atlas and Catalog of
interacting galaxies (Vorontsov-Velyaminov
\cite{Vorontsov59,Vorontsov77}) there are  objects with VV
numbers where an interaction is not so evident.  Many such
objects were classified by Vorontsov-Velyaminov as ``nests'' or
often similar systems called ``chains'' or ``pairs in
contact''. The original idea of the nature of such galaxies was
that they are fragmenting unstable systems. However, observations
showed that VV galaxies present a large diversity of types,
with no evidence of instability (at least in most cases).
It is clear now that the objects in question include both
galaxies of strange and unusual shapes (either single objects or
mergers), and multiple systems as well, in which only detailed
studies allow to disentangle separate components. In
particular, the study of ionized or neutral gas velocity
distributions may help to resolve the enigma of a galaxy with
strange morphology.

The complimentary spectral data are also useful to better constrain the
nature of the studied galaxies. In particular, chemical element abundances in
regions of current starburst and ages of recent starbursts are important to
understand  the processes responsible for their peculiarities.

For some  of these galaxies H$\alpha$-line kinematic data
have been obtained with the 6\,m telescope in the beginning of the 1980s
(Arkhipova et al.~\cite{Arkhipova81}, \cite{Arkhipova87a},
\cite{Arkhipova87b}, \cite{Arkhipova87c}) using
the Long-Slit spectrograph with an image tube and photoplate  as a detector.
New observational opportunities based on use of CCD detectors allowed
a higher sensitivity and larger dynamical range data.
This strengthened our interest  more detailed studies of the subsample
of VV galaxies with no traces of interaction.

For the current study we selected about 30 VV galaxies. They look
like tight multiple systems or single clumpy galaxies, in which such
indicators of interaction as well defined tidal tails or bridges
are absent.  In Paper I we published the results of four galaxies --
VV~432, VV~543E, VV~543W and VV~747.
In this paper, the second in the series, we present results of
long-slit spectroscopy with the SAO 6\,m telescope for five more
VV galaxies:
VV~080 (NGC~14), VV~131 (UGC 4874), VV~499 (DDO~053), VV~523 (NGC~3991) and
VV~531 (NGC~1156).

For all galaxies we obtained the velocity distribution
along the slit using the H$\alpha$ line, determined  relative intensities of
detected emission lines, derived abundances of
oxygen and some other elements, and estimated characteristic ages of
starbursts, as seen in the brightest individual knots.
In Section~\ref{observations} we describe observations, data
reduction, and the analysis of obtained observational parameters.
In Section~\ref{Individual_prop} we consider individual properties
of the studied galaxies.
A summary of the obtained results and preliminary conclusions
are presented in Section~\ref{Conclusions}.

\section{Spectral observations, data reduction and analysis}
\label{observations}

\subsection{Observations}

All results presented below were obtained  with the SAO
6\,m telescope during  seven runs in February, April, July, September
and November 1999, October 2000, and December 2001 (see for details
Table~\ref{Tab2}).
The Long-Slit spectrograph (Afanasiev et al.
\cite{Afanasiev95}) at the 6\,m telescope prime focus was equipped with a
Photometrics CCD-detector PM1024 (with a 24$\times$24$\mu$m pixel size)
installed at Schmidt-Cassegrain camera F/1.5.
Observations were conducted mainly using the software package {\tt NICE} in
MIDAS, described by Kniazev \& Shergin (\cite{Kniazev95}).
Long slit spectra (120\arcsec)  for  abundance analysis
were obtained with either a 325 gr.mm$^{-1}$ grating
(4.6~\AA~pix$^{-1}$), or with a 651 gr.mm$^{-1}$ grating
(2.4~\AA~pix$^{-1}$). The slit width was 1.5\arcsec\ and 2.0\arcsec\,
respectively. For the study of ionized gas kinematics
a 1302 gr.mm$^{-1}$ grating (1.2~\AA~pix$^{-1}$) with a slit of
2\arcsec$\times$120\arcsec\ was used.
The scale along the slit was 0.39\arcsec~pix$^{-1}$ for all three set-ups.
Seeing during observations was mostly in the range of 1.5\arcsec\
 to 2.5\arcsec.
The resulting resolution was about 14--15~\AA\ for the first set-up, 7~\AA\
for the second, and about 3.2~\AA\ for the third set-up.
Reference spectra of an He--Ne--Ar lamp were recorded before or after
each observation to provide wavelength calibration.
Spectrophotometric standard stars from Massey et al. (\cite{Massey88})
were observed for flux calibration at least twice a night.

Results of spectroscopy are presented in various forms. We show
in the top panels of Figures 1,3,5,7 and 9 direct
images of the studied galaxies, extracted from the Digitized Sky Survey (DSS)
with the position of the long slit indicated by a bar.
Corresponding 2D spectra are shown in the middle panels of these figures.
Brightness profiles of H$\alpha$ line along the slit and the corresponding
Position--Velocity (P--V) diagrams are presented in Figures 2,4,6,8 and 10.
In bottom panels of Figures 1,3,5,7 and 9
we present 1D spectra (extracted from the above 2D spectra), which
were used for measurements of line intensities and determination of physical
conditions and chemical abundances in these \ion{H}{ii} regions.

\begin{table*}[hbtp]
\centering{
\caption{Line intensities in the brightest knots of VV~080, VV~131 and VV~499}
\label{Tab3}
\begin{tabular}{lcccccc} \hline \hline
\rule{0pt}{10pt}
& \MC{2}{c}{VV~080 (a)}         &  \MC{2}{c}{VV~131 (a)} &  \MC{2}{c}{VV~499 (b)}       \\ \hline
\rule{0pt}{10pt}
$\lambda_{0}$(\AA) Ion          & F($\lambda$)/F(H$\beta$)&I($\lambda$)/I(H$\beta$)     &F($\lambda$)/F(H$\beta$)&I($\lambda$)/I(H$\beta$) &F($\lambda$)/F(H$\beta$)&I($\lambda$)/I(H$\beta$)        \\ \hline
3727\ [O\ {\sc ii}]\            & 4.691$\pm$0.474 & 6.476$\pm$0.692 & 1.394$\pm$0.112 & 1.394$\pm$0.119 & 1.636$\pm$0.131 & 1.803$\pm$0.156 \\
3868\ [Ne\ {\sc iii}]\          & ---             & ---             & 0.799$\pm$0.074 & 0.799$\pm$0.076 & 0.285$\pm$0.037 & 0.310$\pm$0.042 \\
3889\ He\ {\sc i}\ +\ H8\       & ---             & ---             & 0.348$\pm$0.047 & 0.348$\pm$0.064 & 0.229$\pm$0.034 & 0.254$\pm$0.062 \\
3967\ [Ne\ {\sc iii}]\ +\ H7\   & ---             & ---             & 0.538$\pm$0.049 & 0.538$\pm$0.059 & 0.291$\pm$0.032 & 0.319$\pm$0.056 \\
4101\ H$\delta$\                & 0.219$\pm$0.067 & 0.269$\pm$0.115 & 0.307$\pm$0.026 & 0.307$\pm$0.034 & 0.339$\pm$0.035 & 0.364$\pm$0.050 \\
4340\ H$\gamma$\                & 0.461$\pm$0.057 & 0.528$\pm$0.087 & 0.566$\pm$0.042 & 0.566$\pm$0.045 & 0.451$\pm$0.037 & 0.472$\pm$0.045 \\
4363\ [O\ {\sc iii}]\           & ---             & ---             & 0.135$\pm$0.013 & 0.135$\pm$0.014 & 0.083$\pm$0.013 & 0.086$\pm$0.014 \\
4861\ H$\beta$\                 & 1.000$\pm$0.086 & 1.000$\pm$0.095 & 1.000$\pm$0.072 & 1.000$\pm$0.072 & 1.000$\pm$0.074 & 1.000$\pm$0.076 \\
4959\ [O\ {\sc iii}]\           & 0.554$\pm$0.054 & 0.541$\pm$0.053 & 2.108$\pm$0.152 & 2.108$\pm$0.152 & 0.993$\pm$0.073 & 0.984$\pm$0.073 \\
5007\ [O\ {\sc iii}]\           & 1.745$\pm$0.177 & 1.684$\pm$0.173 & 6.041$\pm$0.468 & 6.041$\pm$0.468 & 2.867$\pm$0.270 & 2.830$\pm$0.268 \\
5876\ He\ {\sc i}\              & ---             & ---             & 0.072$\pm$0.007 & 0.072$\pm$0.007 & ---             & ---             \\
6300\ [O\ {\sc i}]\             & 0.191$\pm$0.045 & 0.141$\pm$0.034 & 0.017$\pm$0.004 & 0.017$\pm$0.004 & ---             & ---             \\
6364\ [O\ {\sc i}]\             & 0.091$\pm$0.051 & 0.066$\pm$0.037 & ---             & ---             & ---             & ---             \\
6548\ [N\ {\sc ii}]\            & 0.191$\pm$0.034 & 0.135$\pm$0.025 & 0.014$\pm$0.005 & 0.014$\pm$0.005 & 0.014$\pm$0.054 & 0.013$\pm$0.049 \\
6563\ H$\alpha$\                & 4.086$\pm$0.325 & 2.885$\pm$0.252 & 2.781$\pm$0.198 & 2.781$\pm$0.215 & 3.070$\pm$0.238 & 2.754$\pm$0.233 \\
6584\ [N\ {\sc ii}]\            & 0.572$\pm$0.058 & 0.403$\pm$0.044 & 0.042$\pm$0.008 & 0.042$\pm$0.008 & 0.044$\pm$0.056 & 0.039$\pm$0.051 \\
6678\ He\ {\sc i}\              & 0.073$\pm$0.040 & 0.050$\pm$0.028 & 0.027$\pm$0.004 & 0.027$\pm$0.004 & ---             & ---             \\
6717\ [S\ {\sc ii}]\            & 0.776$\pm$0.071 & 0.534$\pm$0.053 & 0.083$\pm$0.008 & 0.083$\pm$0.009 & 0.185$\pm$0.042 & 0.165$\pm$0.038 \\
6731\ [S\ {\sc ii}]\            & 0.601$\pm$0.059 & 0.413$\pm$0.044 & 0.062$\pm$0.007 & 0.062$\pm$0.007 & 0.113$\pm$0.040 & 0.101$\pm$0.036 \\
7065\ He\ {\sc i}\              & 0.029$\pm$0.016 & 0.019$\pm$0.010 & 0.024$\pm$0.005 & 0.024$\pm$0.005 & ---             & ---             \\
7136\ [Ar\ {\sc iii}]\          & 0.091$\pm$0.024 & 0.059$\pm$0.015 & 0.049$\pm$0.006 & 0.049$\pm$0.006 & ---             & ---             \\
  & & \\
C(H$\beta$)\ dex         & \MC {2}{c}{0.46$\pm$0.10}       & \MC {2}{c}{0.00$\pm$0.09}       & \MC {2}{c}{0.14$\pm$0.10} \\
EW(abs)\ \AA\            & \MC {2}{c}{0.00$\pm$2.18}       & \MC {2}{c}{0.00$\pm$2.96}       & \MC {2}{c}{0.55$\pm$4.55} \\
F(H$\beta$)$^\mathrm{(a)}$\ & \MC {2}{c}{43$\pm$3}            & \MC {2}{c}{118$\pm$6}           & \MC {2}{c}{56$\pm$3}     \\
EW(H$\beta$)\ \AA\       & \MC {2}{c}{55$\pm$3}            & \MC {2}{c}{303$\pm$15}          & \MC {2}{c}{273$\pm$14}    \\
$V_{\rm Hel}$         & \MC {2}{c}{884$\pm$18}          & \MC {2}{c}{2790$\pm$27}         & \MC {2}{c}{-10$\pm$6}      \\
T$_{\rm e}$([OIII]) K          & \MC {2}{c}{---}                 & \MC {2}{c}{16000$\pm$900}       & \MC {2}{c}{18800$\pm$1900}      \\
12+$\log$(O/H)          & \MC {2}{c}{8.6$\pm$0.1$^\mathrm{(b)}$}   & \MC {2}{c}{7.85$\pm$0.06}       & \MC {2}{c}{7.52$\pm$0.08}       \\
$\log$(N/O)            & \MC {2}{c}{---}                 & \MC {2}{c}{-1.62$\pm$0.10}      & \MC {2}{c}{-1.72$\pm$0.44}      \\
$\log$(Ne/O)           & \MC {2}{c}{---}                 & \MC {2}{c}{-0.54$\pm$0.09}      & \MC {2}{c}{-0.65$\pm$0.14}      \\
\hline \hline
\multicolumn{6}{l}{$^\mathrm{(a)}$ -- in units 10$^{-16}$~erg~cm$^{-1}$~s$^{-1}$} \\
\multicolumn{6}{l}{$^\mathrm{(b)}$ -- determined by combination of empirical methods (see section~\ref{VV080abun})} \\
\end{tabular}
 }
\end{table*}

\begin{table*}[hbtp]
\centering{
\caption{Line intensities in the brightest knots of VV~523 and VV~531}
\label{Tab4}
\begin{tabular}{lcccc} \hline \hline
\rule{0pt}{10pt}
& \MC{2}{c}{VV~523 (d)}     &  \MC{2}{c}{VV~531 (e)}       \\ \hline
\rule{0pt}{10pt}
$\lambda_{0}$(\AA) Ion         & F($\lambda$)/F(H$\beta$)&I($\lambda$)/I(H$\beta$)     &F($\lambda$)/F(H$\beta$)&I($\lambda$)/I(H$\beta$)        \\ \hline
3727\ [O\ {\sc ii}]\           & 0.953$\pm$0.087 & 0.847$\pm$0.092 & 2.309$\pm$0.171 & 3.093$\pm$0.252 \\
3868\ [Ne\ {\sc iii}]\         & 0.205$\pm$0.032 & 0.182$\pm$0.033 & 0.274$\pm$0.025 & 0.351$\pm$0.035 \\
3889\ He\ {\sc i}\ +\ H8\      & ---             & ---             & 0.157$\pm$0.018 & 0.220$\pm$0.035 \\
3967\ [Ne\ {\sc iii}]\ +\ H7\  & 0.131$\pm$0.025 & 0.291$\pm$0.038 & 0.193$\pm$0.018 & 0.258$\pm$0.032 \\
4101\ H$\delta$\               & 0.124$\pm$0.018 & 0.266$\pm$0.029 & 0.221$\pm$0.018 & 0.281$\pm$0.030 \\
4340\ H$\gamma$\               & 0.329$\pm$0.030 & 0.426$\pm$0.036 & 0.394$\pm$0.030 & 0.456$\pm$0.038 \\
4363\ [O\ {\sc iii}]\          & 0.013$\pm$0.010 & 0.012$\pm$0.010 & 0.033$\pm$0.009 & 0.037$\pm$0.010 \\
4471\ He\ {\sc i}\             & 0.028$\pm$0.011 & 0.024$\pm$0.011 & 0.035$\pm$0.006 & 0.038$\pm$0.007 \\
4861\ H$\beta$\                & 1.000$\pm$0.082 & 1.000$\pm$0.085 & 1.000$\pm$0.076 & 1.000$\pm$0.078 \\
4959\ [O\ {\sc iii}]\          & 1.087$\pm$0.087 & 0.966$\pm$0.087 & 1.298$\pm$0.156 & 1.258$\pm$0.154 \\
5007\ [O\ {\sc iii}]\          & 3.220$\pm$0.290 & 2.863$\pm$0.290 & 3.982$\pm$0.329 & 3.818$\pm$0.321 \\
5876\ He\ {\sc i}\             & 0.117$\pm$0.018 & 0.104$\pm$0.018 & 0.135$\pm$0.011 & 0.109$\pm$0.009 \\
6300\ [O\ {\sc i}\             & ---             & ---             & 0.054$\pm$0.008 & 0.040$\pm$0.006 \\
6312\ [S\ {\sc iii}]\          & ---             & ---             & 0.019$\pm$0.006 & 0.014$\pm$0.005 \\
6364\ [O\ {\sc i}]\            & 0.014$\pm$0.006 & 0.012$\pm$0.006 & 0.017$\pm$0.005 & 0.012$\pm$0.004 \\
6548\ [N\ {\sc ii}]\           & 0.079$\pm$0.008 & 0.071$\pm$0.008 & 0.104$\pm$0.009 & 0.075$\pm$0.007 \\
6563\ H$\alpha$\               & 2.961$\pm$0.228 & 2.684$\pm$0.248 & 4.000$\pm$0.294 & 2.869$\pm$0.233 \\
6584\ [N\ {\sc ii}]\           & 0.238$\pm$0.021 & 0.212$\pm$0.023 & 0.312$\pm$0.026 & 0.223$\pm$0.020 \\
6678\ He\ {\sc i}\             & 0.029$\pm$0.005 & 0.026$\pm$0.005 & 0.050$\pm$0.006 & 0.035$\pm$0.005 \\
6717\ [S\ {\sc ii}]\           & 0.260$\pm$0.021 & 0.231$\pm$0.023 & 0.333$\pm$0.025 & 0.233$\pm$0.020 \\
6731\ [S\ {\sc ii}]\           & 0.196$\pm$0.016 & 0.174$\pm$0.018 & 0.256$\pm$0.020 & 0.178$\pm$0.015 \\
7065\ He\ {\sc i}\             & 0.025$\pm$0.005 & 0.022$\pm$0.005 & 0.031$\pm$0.006 & 0.021$\pm$0.004 \\
7136\ [Ar\ {\sc iii}]\         & 0.114$\pm$0.011 & 0.102$\pm$0.011 & 0.118$\pm$0.011 & 0.078$\pm$0.008 \\
  & & \\
C(H$\beta$)\ dex         & \MC {2}{c}{0.00$\pm$0.10} & \MC {2}{c}{0.42$\pm$0.10}  \\
EW(abs)\ \AA\            & \MC {2}{c}{6.00$\pm$0.58} & \MC {2}{c}{1.50$\pm$1.69}  \\
F(H$\beta$)$^\mathrm{(a)}$ & \MC {2}{c}{757$\pm$44}    & \MC {2}{c}{423$\pm$23}     \\
EW(H$\beta$)\ \AA\       & \MC {2}{c}{48$\pm$3}      & \MC {2}{c}{168$\pm$9}      \\
$V_{\rm Hel}$         & \MC {2}{c}{3355$\pm$24}   & \MC {2}{c}{321$\pm$21} \\
T$_{\rm e}$([OIII]) K          & \MC {2}{c}{8800$\pm$1900} & \MC {2}{c}{11500$\pm$1100}       \\
12+$\log$(O/H)         & \MC {2}{c}{8.65$\pm$0.2$^\mathrm{(b)}$} & \MC {2}{c}{8.17$\pm$0.09}        \\
$\log$(N/O)            & \MC {2}{c}{---}                 & \MC {2}{c}{-1.34$\pm$0.13}       \\
$\log$(Ne/O)           & \MC {2}{c}{---}                 & \MC {2}{c}{-0.63$\pm$0.18}       \\
$\log$(S/O)            & \MC {2}{c}{---}                 & \MC {2}{c}{-1.69$\pm$0.17}       \\
\hline \hline
\multicolumn{4}{l}{$^\mathrm{(a)}$ -- in units 10$^{-16}$~erg~cm$^{-1}$~s$^{-1}$; $^\mathrm{(b)}$ -- combined value, see section~\ref{VV523_abun}} \\
\end{tabular}
 }
\end{table*}

\subsection{Data reduction}

The reduction was performed in  SAO RAS, using various packages of MIDAS
and IRAF. This was described in detail in Paper I, and here we
outline only some differences introduced since that time due to the
development of the methods.


First, the standard steps such as debiasing, flat-fielding and dark noise
subtraction are performed in IRAF for each frame containing the objects and
reference spectra. The removal of cosmic-ray hits was conducted using
MIDAS.
Then 2D linearization of the reference spectra was carried out for each
CCD-line, and was applied to 2D spectra of spectrophotometric standards
and the target galaxies. The original curved object spectra were
straightened. All these procedures were performed in IRAF.

Spectra of standard stars were used to build a spectral response curve
that then was applied to the spectra of target galaxies, taking into
account atmospheric extinction at their different airmasses.
Typical uncertainties of the spectral response curves are $\sim$5\%.
As a result, calibrated 1D spectra of studied galaxies (or their individual
fragments) were obtained by summing several lines of rectified 2D spectra
over the regions of interest.

Finally, to create tables with intensities of emission lines
in 1D spectra, we have integrated fluxes of individual lines and performed
a Gaussian analysis of blended (at our spectral resolution) emission lines,
such as,  H$\gamma$ and [\ion{O}{iii}] $\lambda$4363, or H$\alpha$ and
[\ion{N}{ii}] $\lambda$6548,6584.
Gaussian fitting of individual lines was performed and their centers were
determined in order to measure the redshifts of the objects.
The background level was drawn, either using the algorithm described by
Shergin et al. (\cite{Shergin96}) or manually.
Oxygen abundances of studied galaxies, derived as explained in Sect.
\ref{Abund_determ}, are given along with other global
parameters  in Table \ref{Tab1}.

%

The resulting observed emission line intensities $F$($\lambda$) of various
ions relative to that of H$\beta$, and those corrected for interstellar
extinction and underlying stellar absorption $I$($\lambda$) (according to
the procedure described by Izotov et al.~\cite{Izotov97}),
are presented in Tables~\ref{Tab3} and \ref{Tab4}. The derived extinction
coefficient C(H$\beta$), the equivalent width of absorption Balmer
hydrogen lines EW(abs), equivalent width of emission H$\beta$ line
-- $EW$(H$\beta$) and the observed H$\beta$ flux are shown as well. All data
in these tables are given for the brightest knots of
each studied galaxy. Their names, according to nomenclature presented in
the figures showing the positions of the long slit, follow the galaxy names in
the table headers.

\subsection{Physical conditions and abundance determinations}
\label{Abund_determ}

To analyse emission-line spectra of studied galaxies we followed
the prescriptions given by Aller~(\cite{Aller84}). We choose the line
intensities ratio of [\ion{O}{iii}] $\lambda$(5007+4959)/$\lambda$4363
as the optimal for BCG/\ion{H}{ii}-galaxies in order to determine the
temperature of the \ion{H}{ii} region.
To correct line intensities for internal and external extinction
and underlying Balmer-line absorptions in the stellar continuum, we apply
the self-consistent scheme described by Izotov et al. (\cite{Izotov94}).
To take into account different electron temperatures in the regions of
predominant [\ion{O}{iii}] emission and [\ion{O}{ii}] emission we also
follow the scheme described in detail by Izotov et al. (\cite{Izotov94},
\cite{Izotov97}).

Electron temperatures  in the brightest \ion{H}{ii} regions  of the
observed  VV galaxies and their chemical abundances
are summarized in Tables~\ref{Tab3} and \ref{Tab4}.
Electron densities were determined for all \ion{H}{ii} regions (using the
density sensitive line ratio of the [\ion{S}{ii}] $\lambda$6717,6731 doublet)
in the range of 80 to 160
cm$^{-3}$, with uncertainties of the order of 100\%. So, only  an upper
limit can be inferred: N$_\mathrm{e} \lesssim 150-300$~cm$^{-3}$.

In the case of weakness/absence of [\ion{O}{iii}]-line $\lambda$4363
we used  various empirical methods
suggested, e.g., by McGaugh (\cite{McGaugh91}), or Pilyugin
(\cite{Pilyugin01}), based on the analysis of strong oxygen line intensities.
This relates to VV~080 and VV~523 (see for details the description of
individual galaxy data).
The methods based on relative
intensity of [\ion{N}{ii}] $\lambda$6548,6584 to that of H$\alpha$ (van Zee et
al.~\cite{vanZee98} and Denicolo et al.~\cite{Denicolo02}) were used as well.
Note that Pilyugin (\cite{Pilyugin01}) shows that his  P-method allows one to
get the most precise estimate of O/H.

The important parameter, directly related to equivalent width
of emission line H$\beta$, is the age of the current starburst. It can
be determined through the relationships given in Starburst99 (Leitherer et al.
\cite{Starburst99}).
We used the instantaneous starburst model with the
Salpeter initial mass function (IMF) and $M_{\rm low}$ and $M_{\rm up}$
of 1~$M$\sunn\ and 100~$M$\sunn, respectively.

\subsection{Position--Velocity diagrams}

The procedure of constructing a Position--Velocity (P--V) diagram based on
observations in H$\alpha$ was described in detail in Paper I.
In a further analysis we used only those points in
the P--V diagrams what satisfy the criteria S/N $>$ 3 and
$\sigma_{\rm V}<$~15~\kms.
To check the quality of our method and to estimate the residual trends
left after the application of this restriction, we
apply the same procedure to the two night sky lines in the observed spectrum:
\ion{O}{i} $\lambda$6300.27~\AA\ and $\lambda$6863.96~\AA.
Being transferred to relative velocity, the r.m.s. of scattering
along the slit for these control lines was found to be 1.2~\kms.
The residual systematic trend for corrected night sky lines has
a full range of about 20~\kms\ on the scale of 80\arcsec.
Such accuracy of corrected observed wavelength allows us to
study irregularities in P--V diagrams with amplitudes as low as
10~\kms\ on the angular scale down to 20\arcsec.

\begin{table}[hbtp]
\centering{
\caption{\label{Tab5} EW(H$\beta$) and starburst ages}
\begin{tabular}{cccc} \hline\hline
\rule{0pt}{10pt}
Object & Knot & EW(H$\beta$) & Age (Myr) \\
 (1)   & (2)  &  (3)         &   (4)     \\ \hline
VV~080 & a   &   55$\pm$3    &      5.0   \\
       & c   &   18$\pm$2    &      6.2   \\
       & d   &  136$\pm$7    &      3.1   \\
VV~131 & a   &  303$\pm$15   &      2.9   \\
       & b   &   77$\pm$14   &      5.3   \\
       & c   &   30$\pm$5    &      8.7   \\
VV~499 & a   &   74$\pm$5    &      4.9   \\
       & b   &  273$\pm$14   &      3.2   \\
       & d   &   44$\pm$3    &      7.7   \\
VV~523 & a   &   70$\pm$4    &      4.8   \\
       & b   &   34$\pm$2    &      5.7   \\
       & c   &   33$\pm$3    &      5.7   \\
       & d   &   48$\pm$3    &      5.1   \\
VV~531 & a   &  101$\pm$5    &      4.6   \\
       & b   &   15$\pm$2    &      7.3   \\
       & c   &   22$\pm$4    &      6.6   \\
       & e   &  168$\pm$9    &      4.0   \\
\hline\hline
\end{tabular}
}
\end{table}


\section{Properties of individual galaxies}
\label{Individual_prop}

We summarize the main parameters of the five VV galaxies discussed below
in Table~\ref{Tab1}.
They include the name of the object, coordinates for  epoch J2000,
apparent blue magnitude,
radial heliocentric velocity based on \ion{H}{i} data taken from the
literature with its r.m.s. uncertainty, angular sizes on the isophote
$\mu_{\rm B}$=25 mag. arcsec$^{-2}$, and respective linear size and axial
ratio, absolute blue magnitude and oxygen abundance (12+$\log$(O/H)).
We also present integrated \ion{H}{i} flux, 21-cm line width, M(HI) in solar
units, and the ratio M(HI)/$L_{\rm B}$, in solar units.
All distance-dependent parameters in Table~\ref{Tab1} are derived for the
distances accepted in this table. Absolute magnitudes $M_{\rm B}^{0}$ are
corrected for foreground extinction $A_{\rm B}$, accepted from NED.
No correction for internal extinction was applied.

For galaxies VV~499 and VV~531, distances are determined from
measurements of found resolved stars.
We accept respective values from literature (see
notes to Table~\ref{Tab1}). For more distant galaxies we use their values of
radial velocity
$V_{\rm Hel}$ presented in Table~\ref{Tab1}.
To derive distances, their radial velocities were corrected for the Sun's
motion relative to the Local Group (LG) centroid, according to NED
(Karachentsev \& Makarov \cite{KM96}) and the motion of the LG to the Virgo
cluster with a peculiar velocity of 250~\kms\ (Huchra \cite{Huchra88}; Klypin
et al. \cite{Klypin01}). The Hubble constant is accepted as
75~\kms~Mpc$^{-1}$.

The mass of neutral hydrogen $M_{\rm HI}$ in Table~\ref{Tab1} is derived
directly from the value of integrated flux in the line of 21~cm $F$(HI)
with the use of the standard formula:
\begin{equation}
M(HI) = 2.36\cdot 10^5\cdot F(HI)\times D^2
\end{equation}
where $M$(HI) is in $M$\sunn, $F$(HI) -- in Jy$\cdot$\kms, and
distance $D$ -- in Mpc.

\subsection{VV~080 = NGC~14 = Arp~235}

\subsubsection{General characteristics}

VV~080 is a nearby galaxy with a rather
asymmetric outer dumbell-like boundary (which makes it resemble an
interacting double system on overexposed images). It has a smooth
brightness distribution of outer parts, and  a clumpy bright inner
structure. With its blue luminosity, this galaxy is an
intermediate between normal and dwarf irregulars.
The earlier data on its P--V diagram along the  major axis gave
evidence that this is a single object (Arkhipova et al.~\cite{Arkhipova87b}).
Its \ion{H}{ii} regions (star formation sites) are concentrated in
the inner part of the galaxy with a diameter of $\sim$40\arcsec\ (2.5
kpc).
VV~080 is  not very rich in neutral hydrogen: from the adopted
distance and integrated \ion{H}{i} flux (Table~\ref{Tab1}) it follows that
$M$(HI)$\sim$3.9$\cdot$10$^8~M$\sunn, or
$M$(HI)/$L_{\rm B} \sim$ 0.16 (in solar units).

\subsubsection{Spectroscopy}
\label{VV080abun}

\begin{figure}
   \centering
   \includegraphics[angle=0,width=8.0cm,bb=20 105 580 690]{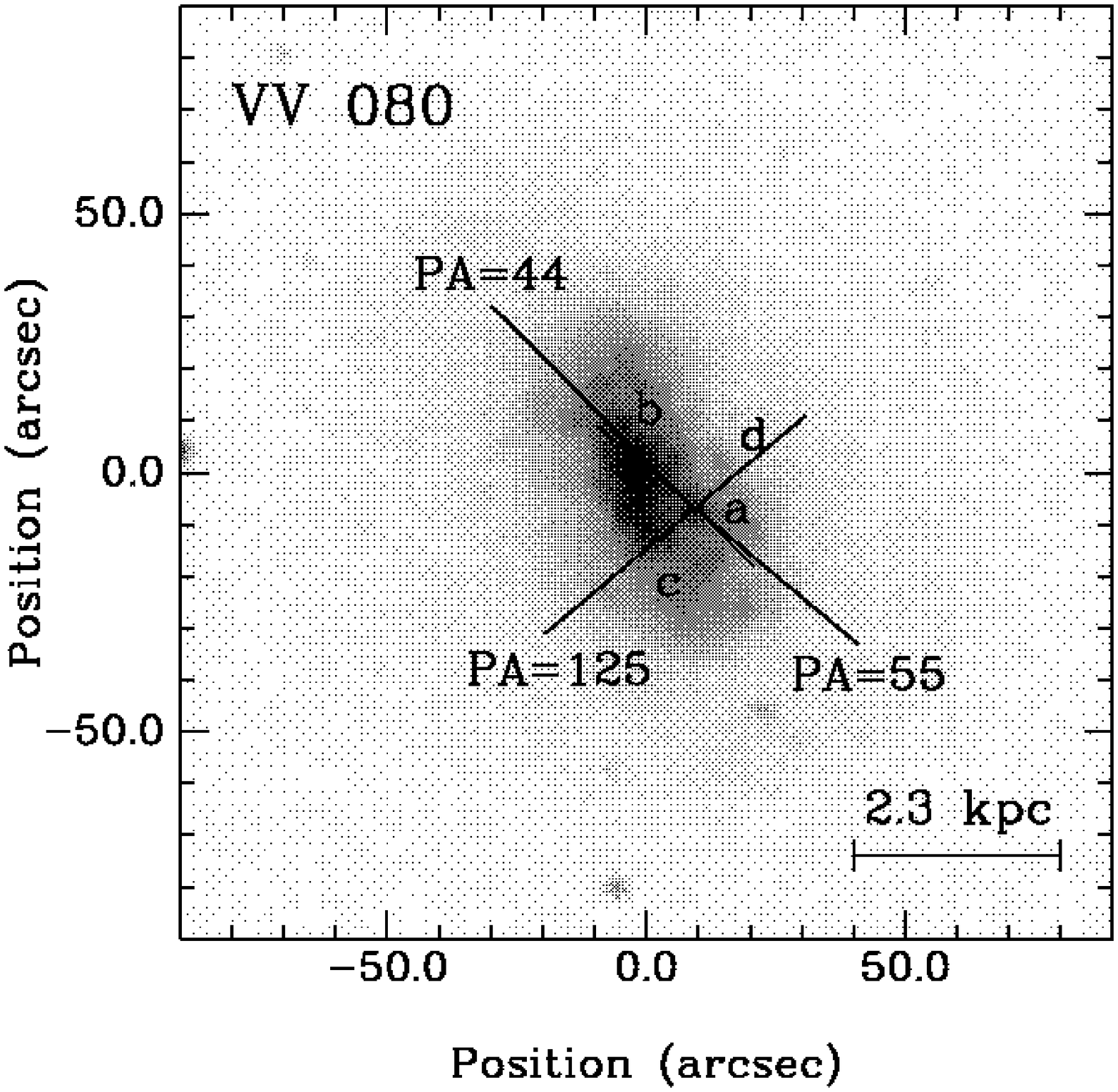}
   \includegraphics[angle=0,width=8.0cm,bb=20 170 580 590]{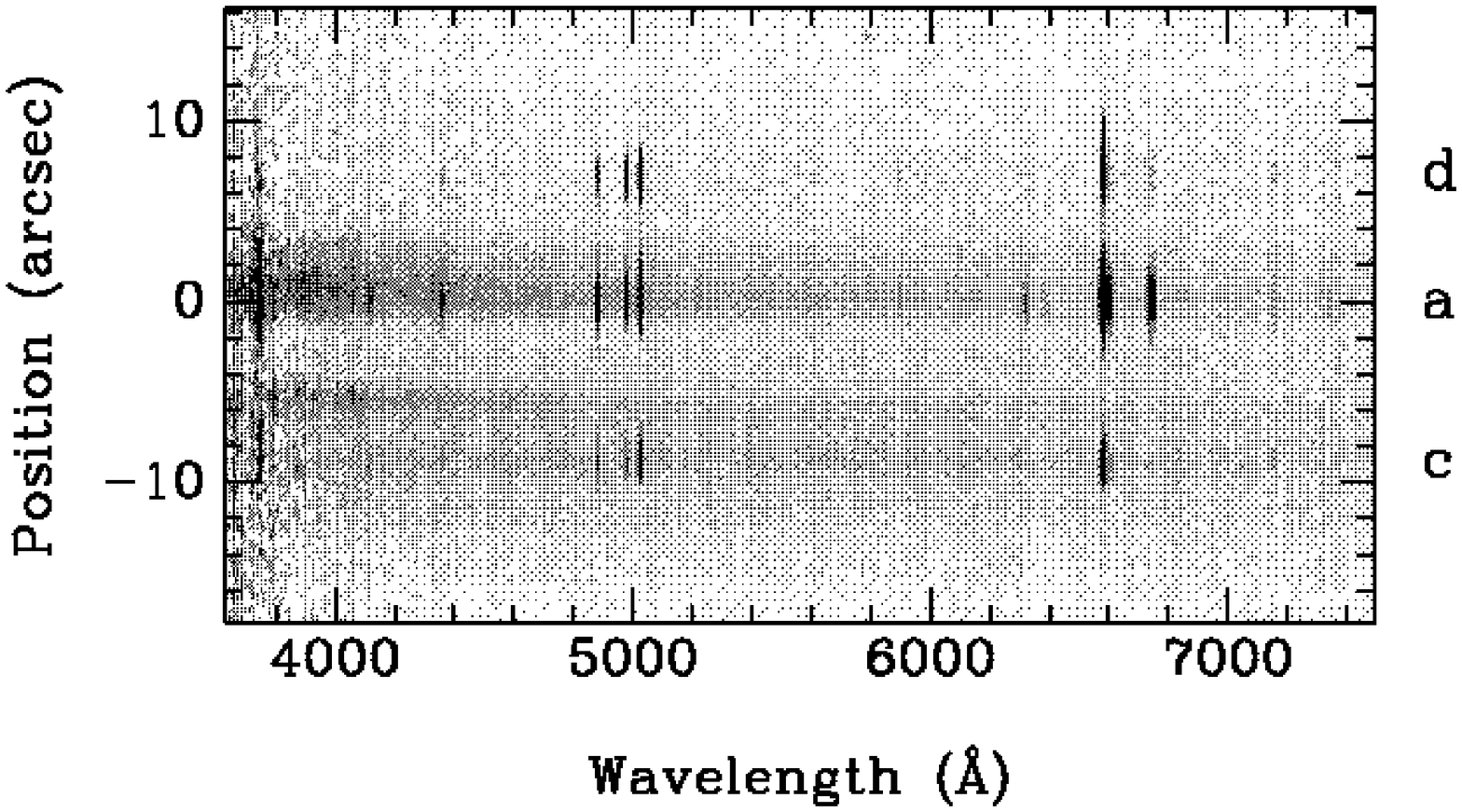}
   \includegraphics[angle=0,width=8.0cm,bb=20 165 580 540]{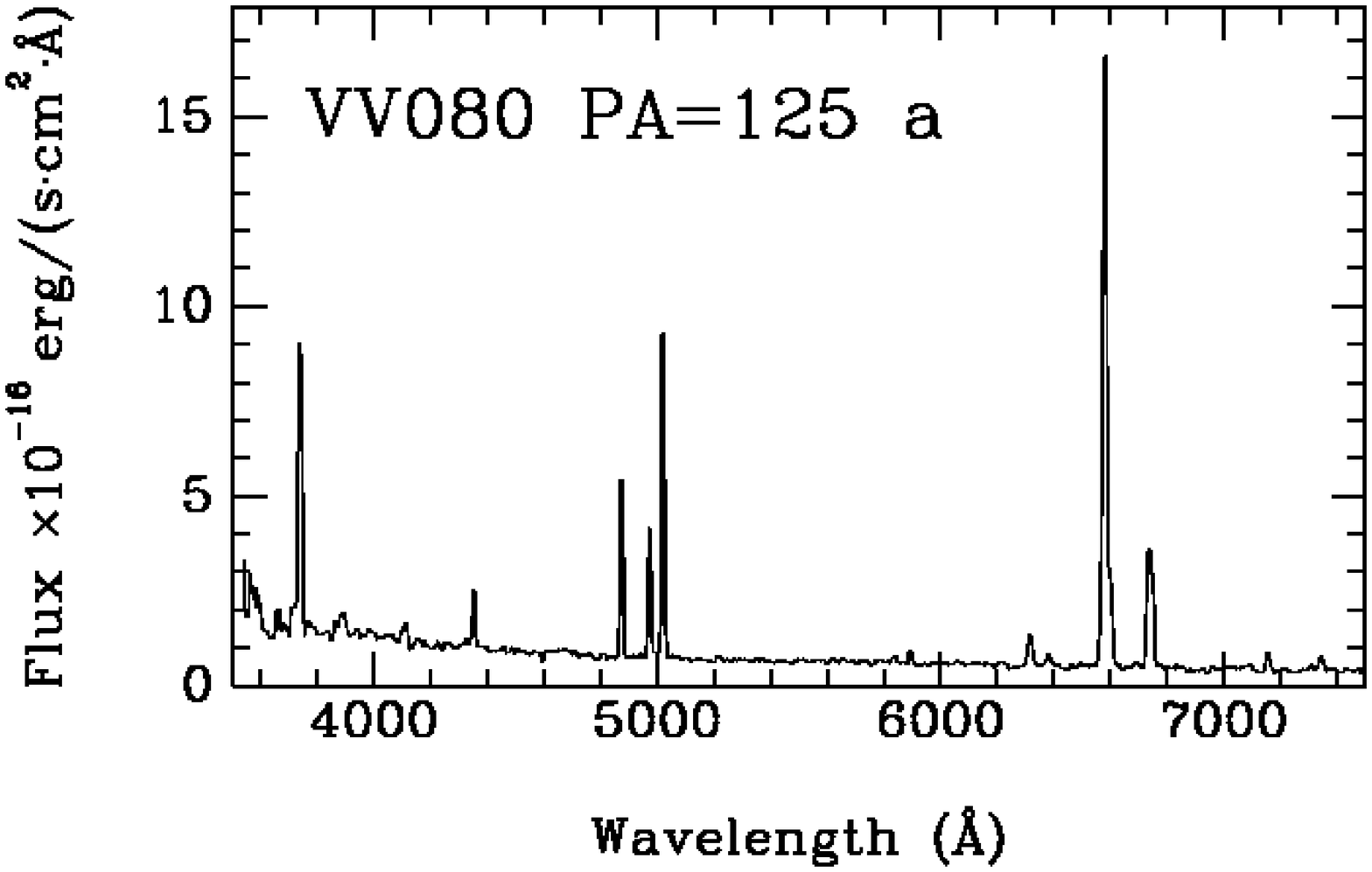}
   \caption{{\it Top:} Blue DSS-2 image of VV~080 with the position of
     long slit  superimposed.
     {\it Middle:} 2-D spectrum of VV~080 for $PA=125$\degr.
     {\it Bottom:} 1-D spectrum of the brightest knot ``a''. }
     \label{VV080fig1}
   \end{figure}

Slit orientations (see Table~\ref{Tab2}) were chosen
in order to cross the most prominent \ion{H}{ii} condensations.
Relative line intensities are presented in Table~\ref{Tab3}
for the brightest knot ``a''.
The  2D spectrum in the slit position with  $PA$=125\degr\ and the 1D
spectrum of knot ``a'' are shown in  Fig.~\ref{VV080fig1}.
Its line ratios  reveal a significant extinction, corresponding
to $C(H\beta)$ = 0.46 (see Table~\ref{Tab3}).

Unfortunately, the non-detection of the [\ion{O}{iii}] $\lambda$4363 line
precludes a direct estimate of O/H in ionized gas of this galaxy.
The parameter $P$ (which is the ratio of I(4959+5007)/(I(4959+5007)+I(3727)),
used by Pilyugin (see section \ref{Abund_determ}) for this galaxy
is 0.26, which is outside
the range (0.4,1.0), for which Pilyugin's relationships for the upper branch
are applicable.
Therefore we do not use that method for this galaxy, and compare the values
of 12+$\log$(O/H),
derived by the other three methods. They give respectively 8.62 (for
intermediate value of ionization parameter) for the McGaugh method, 8.50 for
the Denicolo et al. method, and 8.62 for the van Zee et al. method.
We accept 12+$\log$(O/H)=8.6 as the value most consistent
with all three estimates. Ages of current starbursts in knots
``a'', ``c'' and ``d'', marked on the image and 2D spectrum, are in the
range of $3-6$ Myr (Table~\ref{Tab5}).

\subsubsection{Morphology and gas kinematics}

\begin{figure}
   \centering
   \includegraphics[angle=0,width=8cm,bb=28 105 560 670]{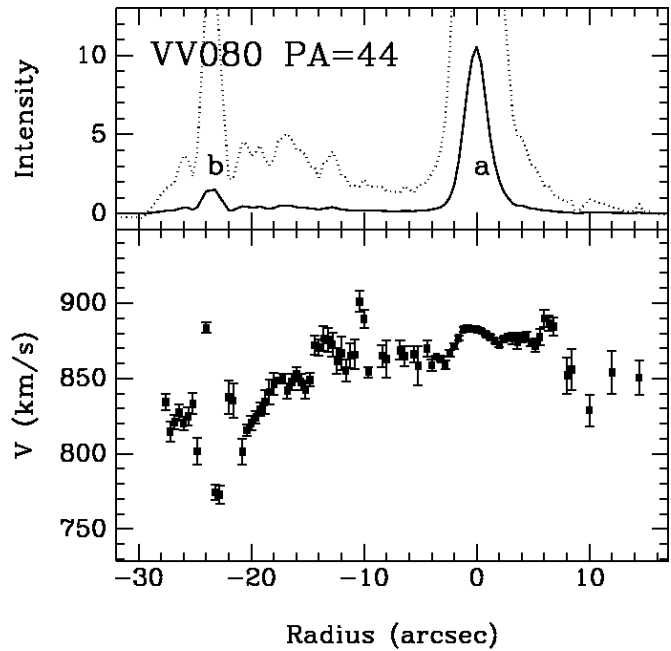}
   \caption[c]{Intensity profile of H$\alpha$ for VV~080
     along the slit with $PA=44$\degr\ and respective P--V diagram.}
     \label{VV080fig2}
   \end{figure}

$PA =$ 44\degr\ of the spectral cut, crossing  knots ``a'' and ``b'',
differs by $\sim$30\degr\ from the $PA$ of the galaxy major axis. Its
P--V diagram indicates very complex gas motion in the inner 2 kpc region.
The small
velocity gradient is interrupted by a negative `wave' with an amplitude
of $\sim$50~\kms\ in the region of knot ``b'', seemingly revealing gas
outflow (shell) from the region of active star formation (SF).
The width of the \ion{H}{i} 21-cm line $W_{\rm 20}$ of VV~080 exceeds
100~\kms.
The rotation velocity  in Irr galaxies is usually observed to grow
monotonously up to the optical borders of a galaxy. Then one might expect
the mean velocity gradient to be about 10
\kms~kpc$^{-1}$, or only (20--25)~\kms\ across the central emission region
of the galaxy. In this case it would be hardly noticeable in the
obtained spectra (see Fig.~\ref{VV080fig2}). It gives  evidence
that the inner velocity gradient exceeds the mean one along the radius of
this galaxy.

Considering $W_{20}$/2$\sin i$, where $i$$\approx$51\degr\ (LEDA),
as the minimal velocity of rotation at the
optical radius of the galaxy, one may obtain lower limits of the
total mass and mass-to-light ratio within the optical radius of
$\sim$5~kpc: $M_{\rm t}$$\approx$6$\cdot$10$^9~M$\sunn,
$M_{\rm t}/L_{\rm B}$$\approx$2 solar units.
Thus the total hydrogen mass does not exceed 8\% of the dynamical
mass of this galaxy.

\subsubsection{Possible interaction with other galaxies}

VV~080 has several neighbouring galaxies at the projected distances
of less than 350 kpc, which implies that VV~080 belongs to a
galaxy group. The most disturbing one is the SA(S)ab galaxy NGC~7814,
$\sim$1\fm2 brighter (M$_{\rm B}$=-19\fm2), situated at the
projected distance of about 300 kpc. Its radial velocity
differs by $\Delta V$=185~\kms. Hence, the tidal effect of this
galaxy can be sufficiently strong  to trigger the observed SF
burst in VV~080.

Since this paper deals mainly with observational data, there is
no space to discuss in detail the strength of tidal action for
each of the target galaxies.
The respective formulae, their discussion and application to the
problem of star formation (SF) trigger in blue compact galaxies (BCGs) can be
found in the paper by Pustilnik et al. (\cite{Pustilnik01},
P01).  Here we briefly, just as an illustration, discuss the
possible role of tidal interaction for VV~080.

Physical mechanisms of SF ignition, which take place in
this galaxy,  can be different.  Here  we have used the
model suggested by Icke (\cite{Icke85}), which involves
generation of shocks in galaxy gas with their subsequent
dissipation and loss of gas stability. We use in the discussion below the
value of threshold pericenter distance $p_\mathrm{0}$ from P01, for
which the tidal action should generate shocks:
$p_\mathrm{0}$$\approx$$R_{\rm gas}\cdot(8\pi\cdot\mu v/s_\mathrm{0})^{1/3}$,
where $R_{\rm gas}$
is the radius of gas disk of disturbed galaxy, $\mu$ is mass
ratio of disturbing and disturbed galaxies, $v$ - rotation
velocity at radius $R$, $s_\mathrm{0}$ - the ``sound'' speed of gas, or
dispersion of gas velocities, accepted as typical value of
10~\kms. The applicability of such estimates  of $p_\mathrm{0}$ had been
tested earlier on the BCG sample from Taylor et al.
(\cite{Taylor95}), for which faint companions were detected (see
P01).

Since the \ion{H}{i} radius of this galaxy is unknown,  to
estimate $p_\mathrm{0}$ we accepted (as follows from several studies,
e.g. Taylor et al. (\cite{Taylor95}), van Zee et al.
(\cite{vanZee98})) that the \ion{H}{i}-to-optical ($R_{\rm 25}$) radius
ratio is $\gtrsim$4 (here $R_{\rm HI}$=$R_{\rm gas}$ is the galaxy radius,
corresponding to the surface gas density of 0.5~$M$\sunn~kpc$^{-2}$),
that is $R_{\rm gas}$$\sim$20 kpc for VV~080. The rotation
velocity, corrected for inclination, is $v$=$W_{\rm 20}/2\sin i$=73~\kms.
The mass ratio $\mu$ can be
estimated to a first approximation from the luminosity ratio of
NGC~7814 and VV~080, taking into account the brightening
$\Delta B$ of VV~080 due to current SF activity. We accept that
$\Delta B$ is at least 0\fm5. Thus, for $\Delta M_{\rm B}$=1\fm2+0\fm5=1\fm7,
$\mu$$\approx$5. With these values we get
$p_\mathrm{0}$=190 kpc. As emphasized in P01, when checking the
neighbouring galaxies as potential disturbers, which tidally
trigger starbursts, we need to account for significant delay
between the time of  maximal action and the subsequent response
revealing itself  as a gas collapse. This delay can reach hundreds
of Myr. During this time the distance between interacting galaxies
will grow by a characteristic value $\Delta l$ of the order of 100
kpc. Summing the estimated $p_\mathrm{0}$ and expected $\Delta l$ we come
to $\sim$300 kpc, close to the observed projected distance
between VV~080 and NGC~7814.

Of course, the above estimates cannot be considered as  strict
evidence for triggering of SF by the tidal action of NGC~7814 on the
gaseous disk of VV~080, especially if we take into account the
absence of its \ion{H}{i} map. We just want to demonstrate using the
example of VV~080 that a simple scheme of such interaction,
already checked for one type of galaxy with active SF, may be
applied to our galaxies. As we see, in the case of VV~080 this
scheme does not contradict the available data on this object.


   \begin{figure}
   \centering
   \includegraphics[angle=0,width=8.0cm,bb=20 105 580 690]{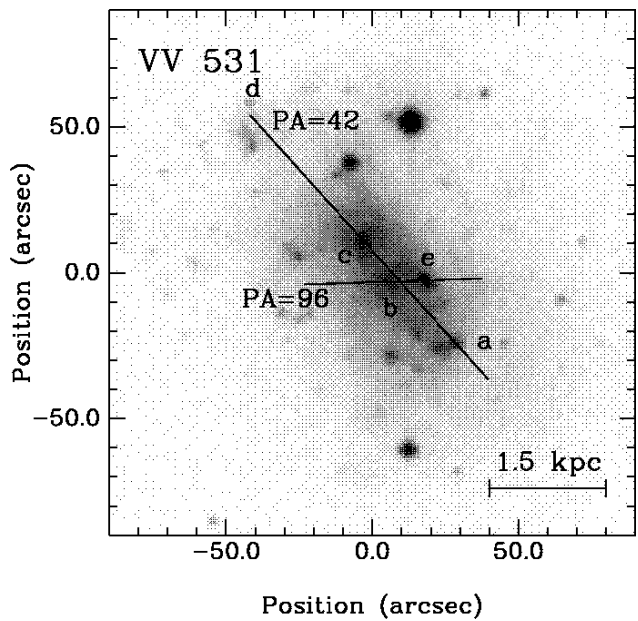}
   \includegraphics[angle=0,width=8.0cm,bb=20 170 580 590]{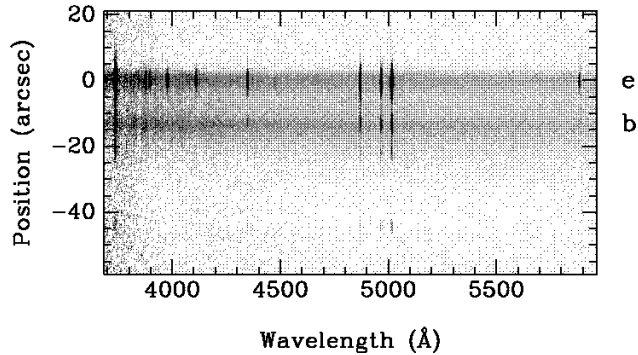}
   \includegraphics[angle=0,width=8.0cm,bb=20 165 580 540]{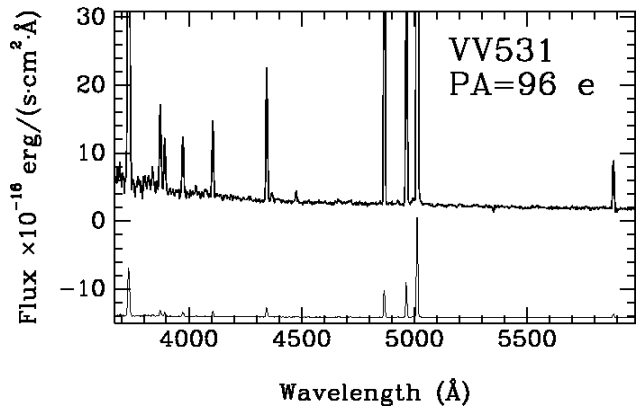}
   \caption[c]{{\it Top:} Blue DSS-2 image of galaxy VV~531 with the positions
	   of the long slit  superimposed.
       {\it Middle:} 2D spectrum of VV~531 along the slit with $PA=96$\degr.
       {\it Bottom:} 1D spectrum of the brightest knot ``e''. }
     \label{VV531fig1}
   \end{figure}

   \begin{figure}
   \centering
   \includegraphics[angle=0,width=8cm,bb=28 135 560 670]{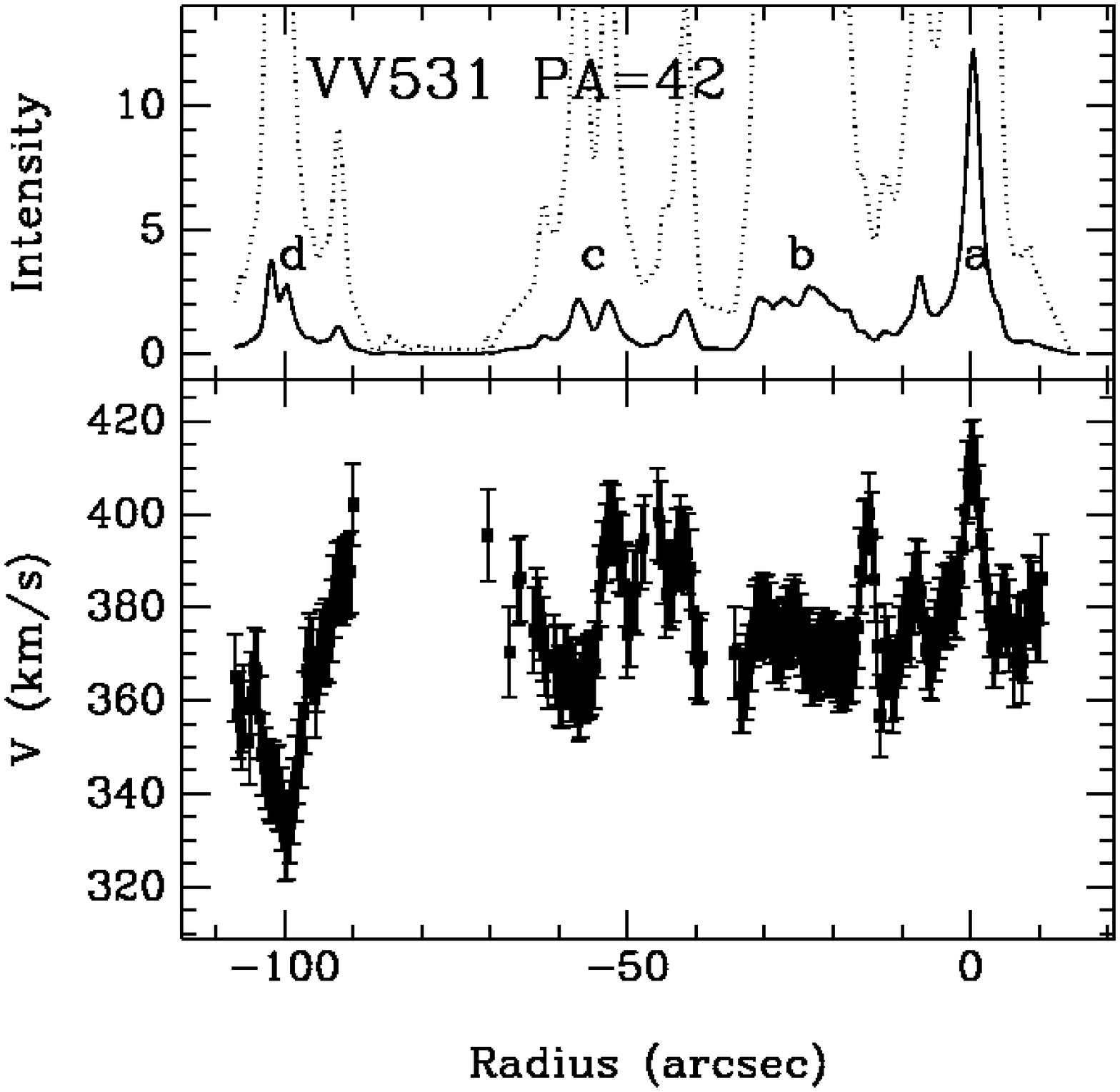}
   \includegraphics[angle=0,width=8cm,bb=28 105 560 670]{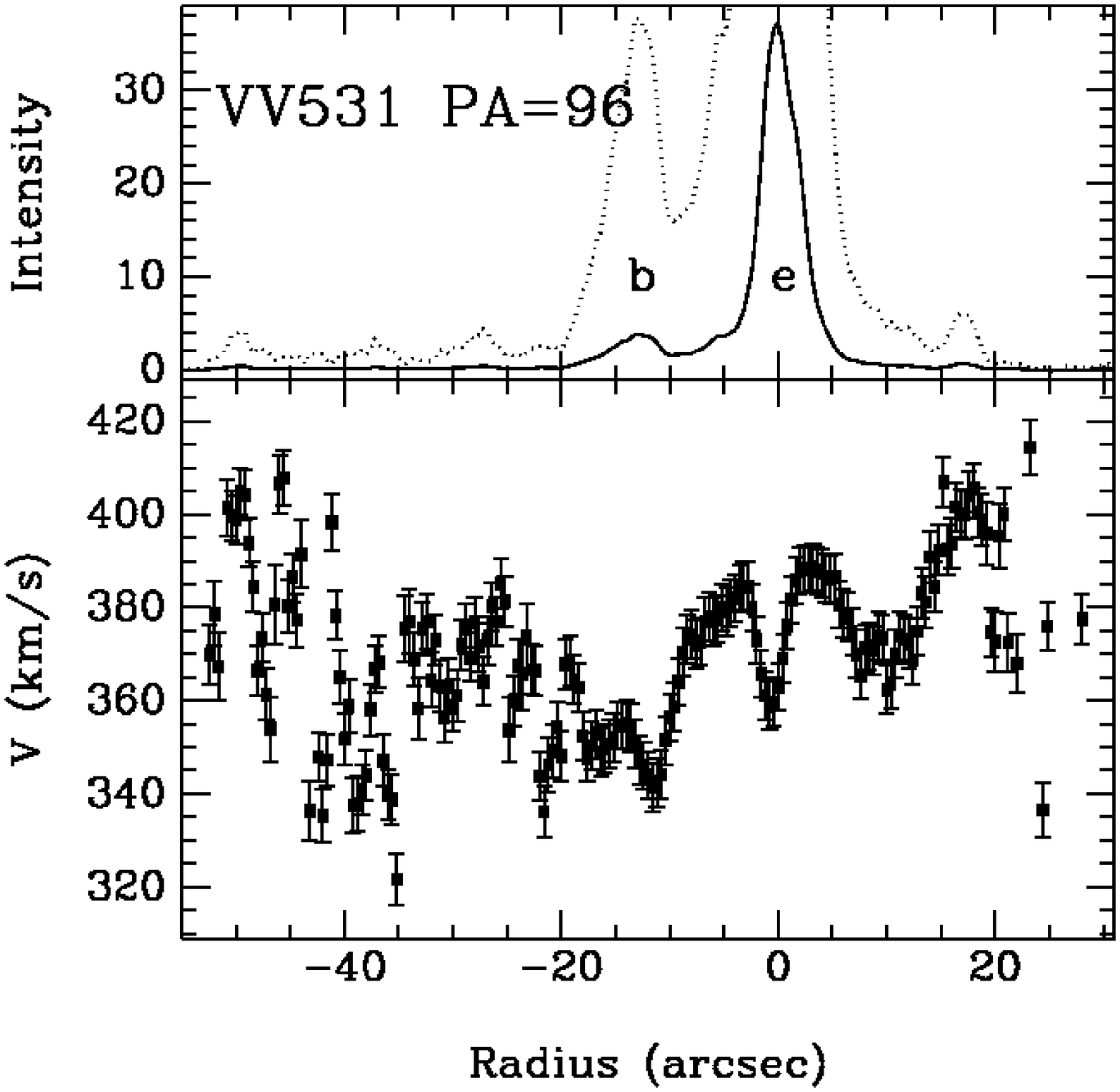}
   \caption[c]{
           Intensity profile of H$\alpha$ for VV~531
           along the slit with $PA=42$\degr\ and 96\degr, and
           respective P--V diagrams.}
     \label{VV531fig2}
   \end{figure}

   \begin{figure}
   \centering
   \includegraphics[angle=0,width=8.0cm,bb=20 105 580 690]{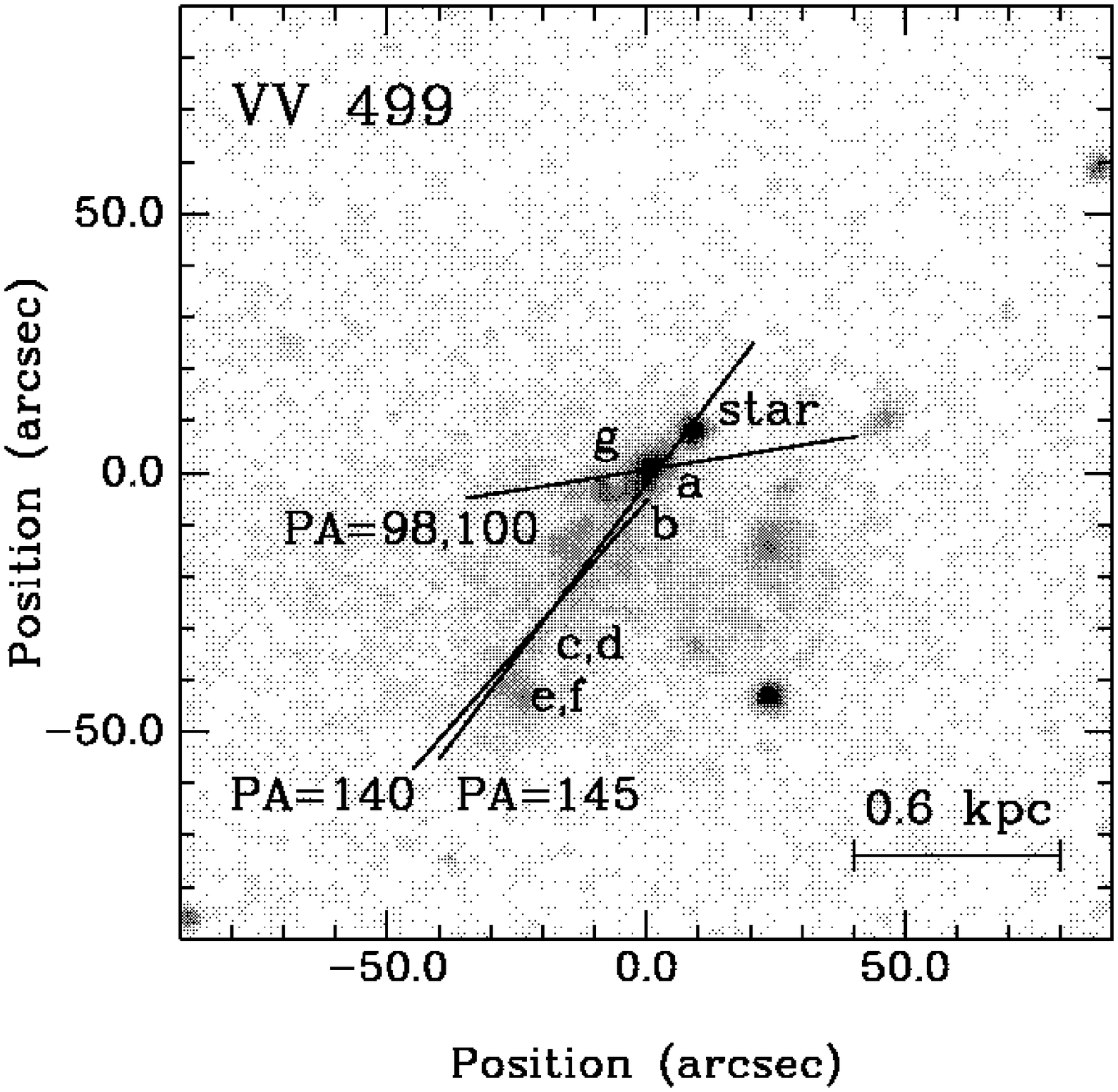}
   \includegraphics[angle=0,width=8.0cm,bb=20 170 580 590]{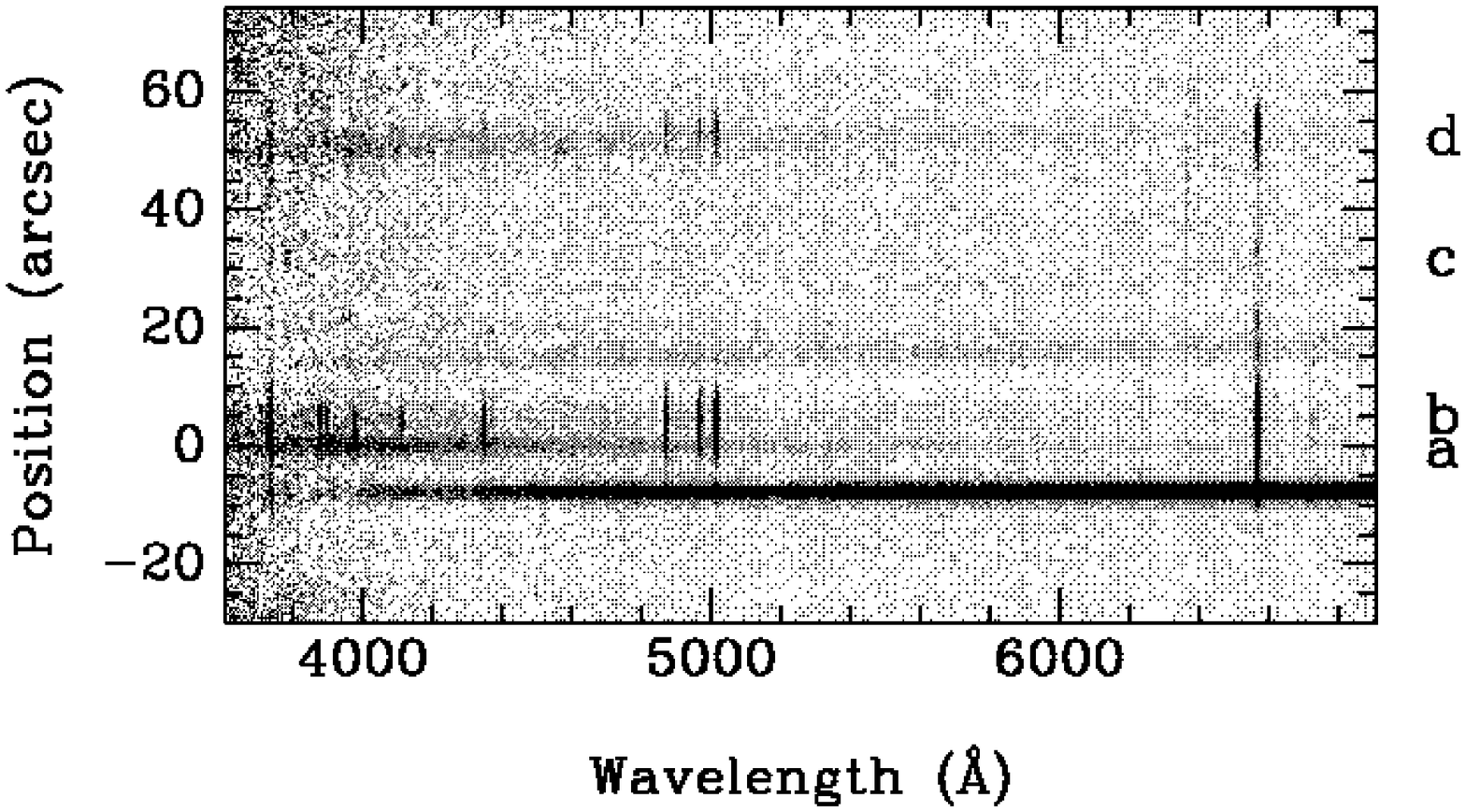}
   \includegraphics[angle=0,width=8.0cm,bb=20 165 580 540]{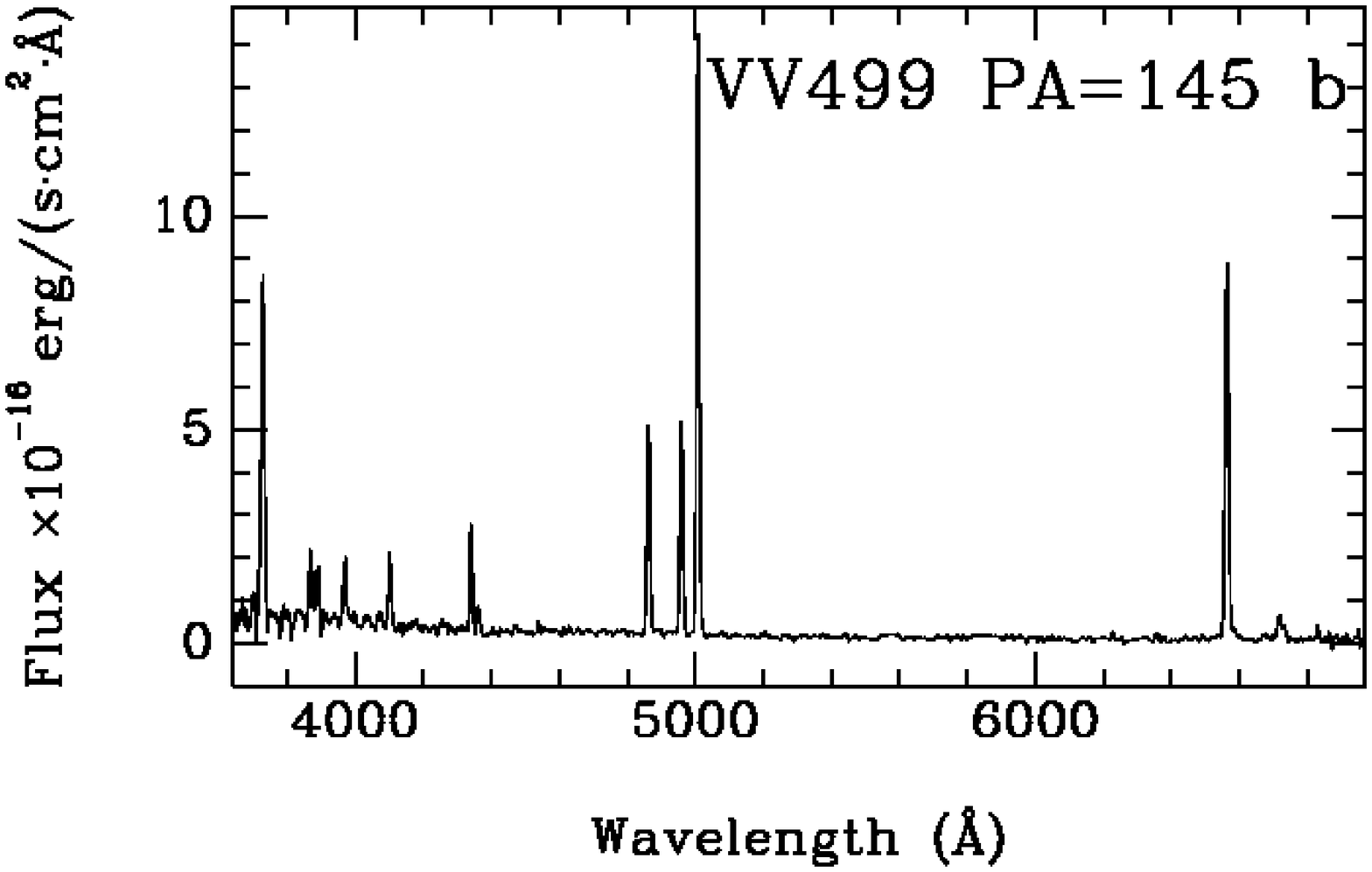}
   \caption[c]{
	  {\it Top:} Blue DSS-2 image of galaxy VV~499 with positions
        of long slit  superimposed.
          {\it Middle:} 2D spectrum of VV~499 along $PA=145$\degr\
          with dispersion 4.6 \AA/pixel.
          {\it Bottom:} 1D spectrum of the brightest knot ``c''.}
     \label{VV499fig1}
   \end{figure}

   \begin{figure}
   \centering
   \includegraphics[angle=0,width=7.5cm,bb=28 135 560 670]{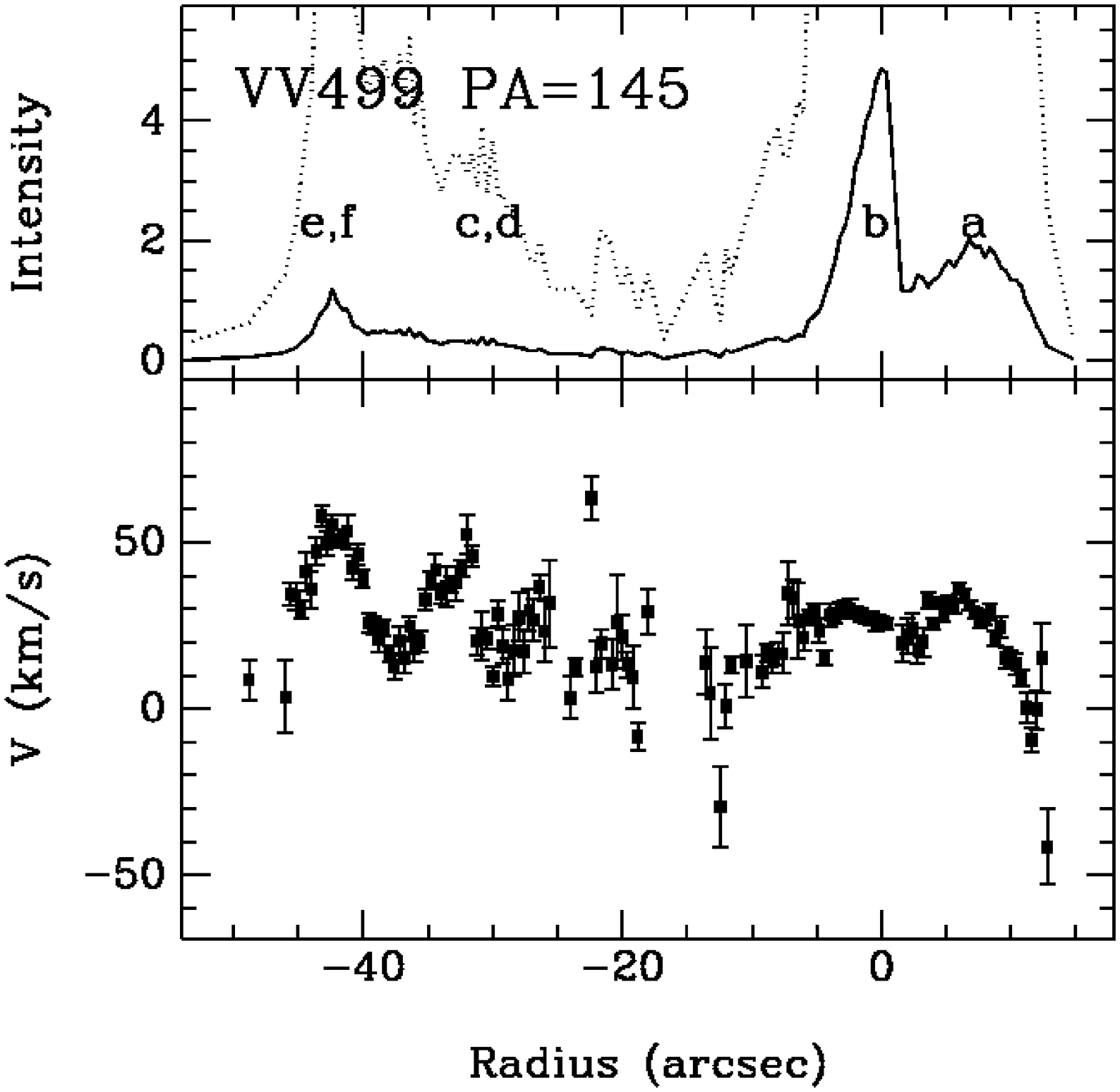}
   \includegraphics[angle=0,width=7.5cm,bb=28 135 560 670]{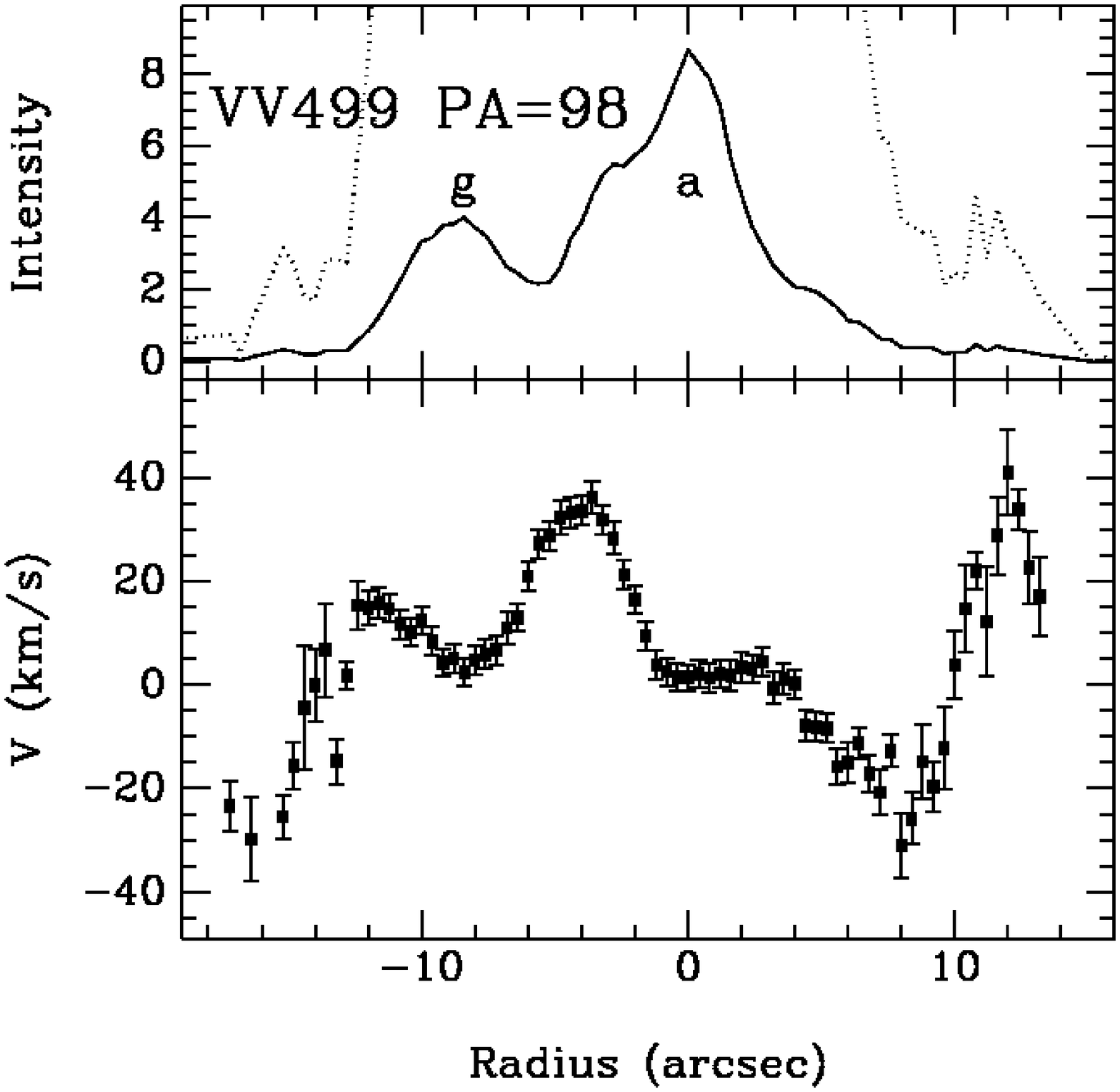}
   \includegraphics[angle=0,width=7.5cm,bb=28 105 560 670]{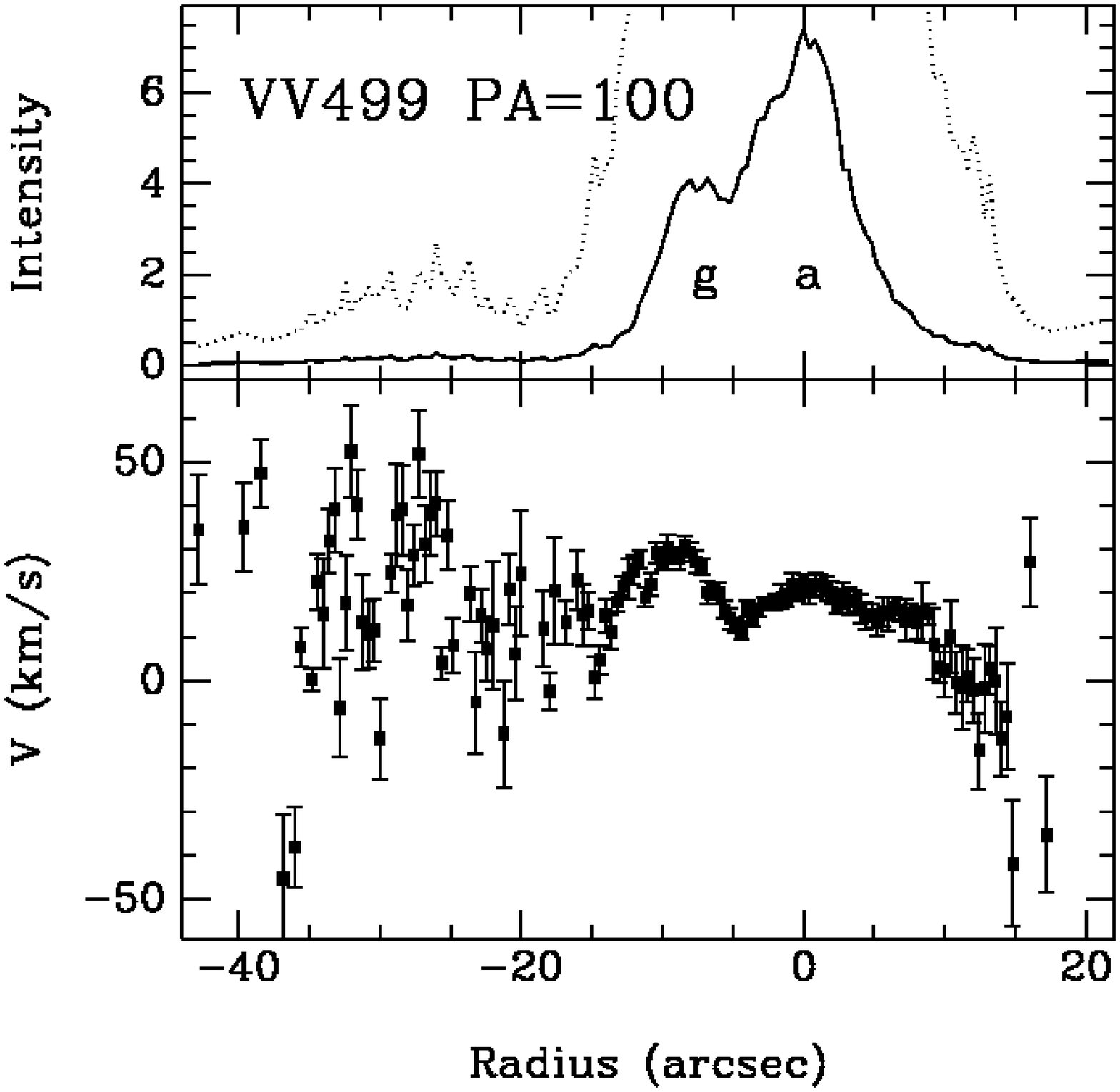}
   \caption[c]{
           Intensity profiles of H$\alpha$ for VV~499
           along the slit with $PA=35$\degr, 82\degr\ and 100\degr,
           and respective P--V diagrams.
          }
     \label{VV499fig2}
   \end{figure}

   \begin{figure}
   \centering
   \includegraphics[angle=0,width=8.0cm,bb=20 105 580 690]{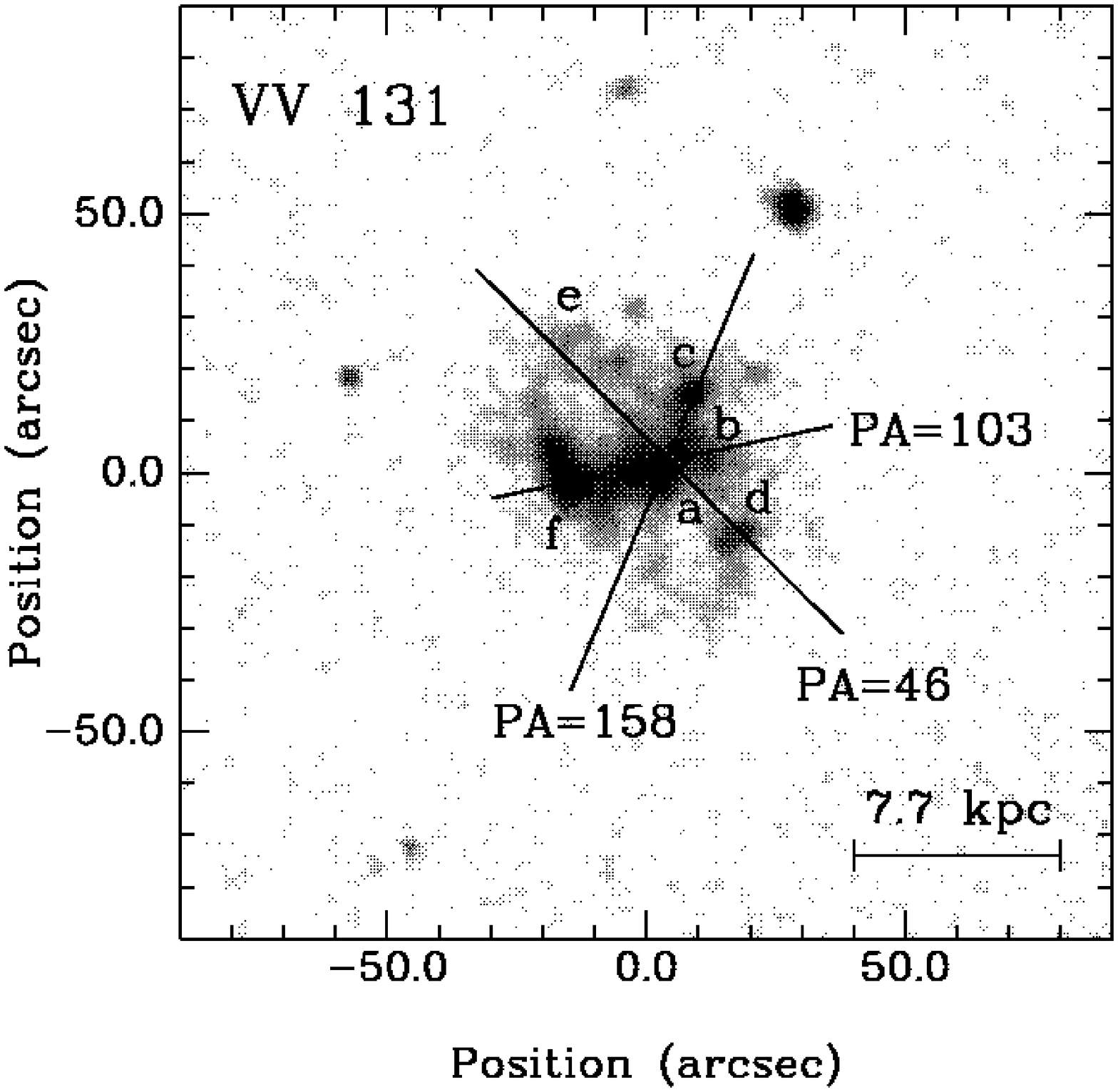}
   \includegraphics[angle=0,width=8.0cm,bb=20 170 580 590]{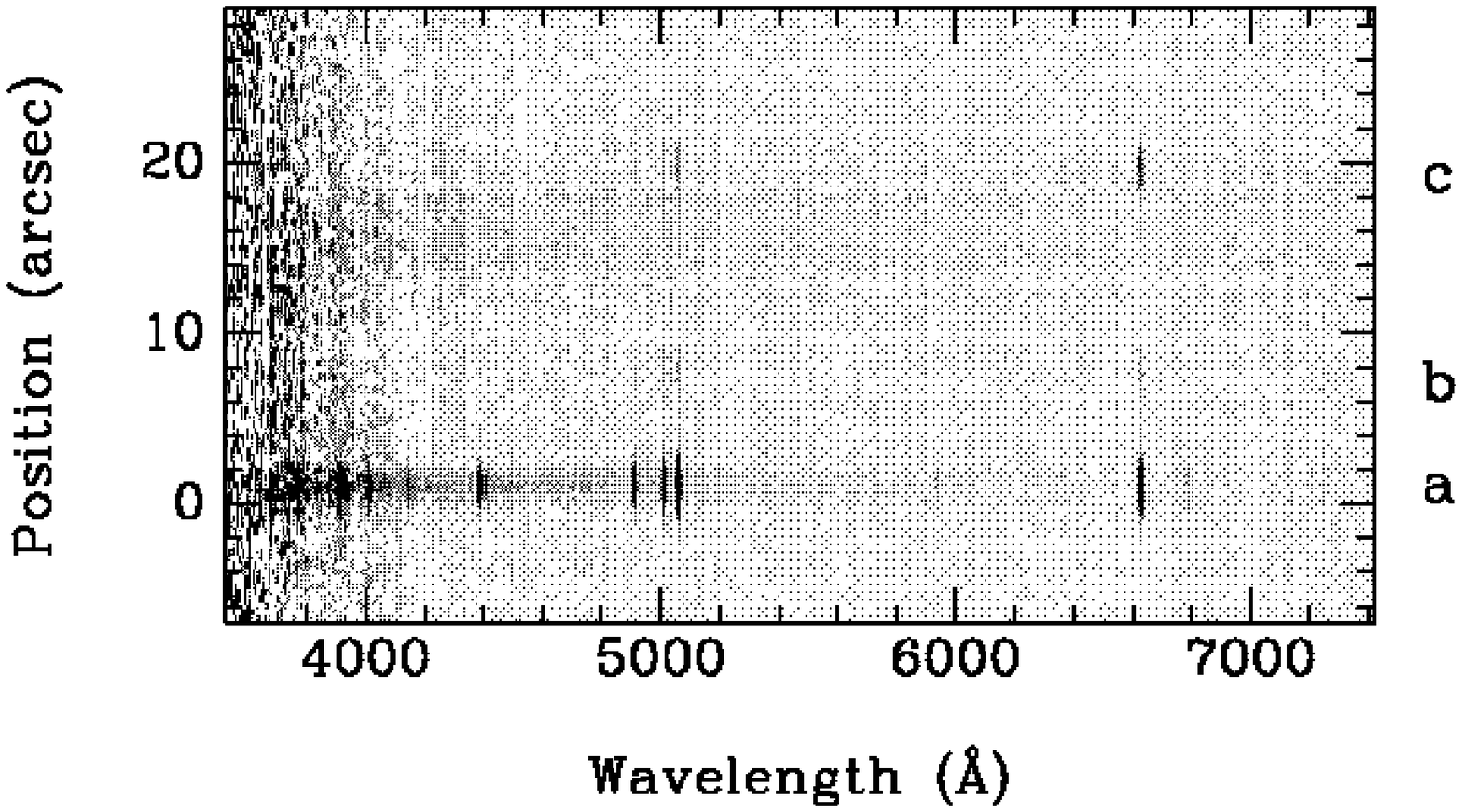}
   \includegraphics[angle=0,width=8.0cm,bb=20 165 580 540]{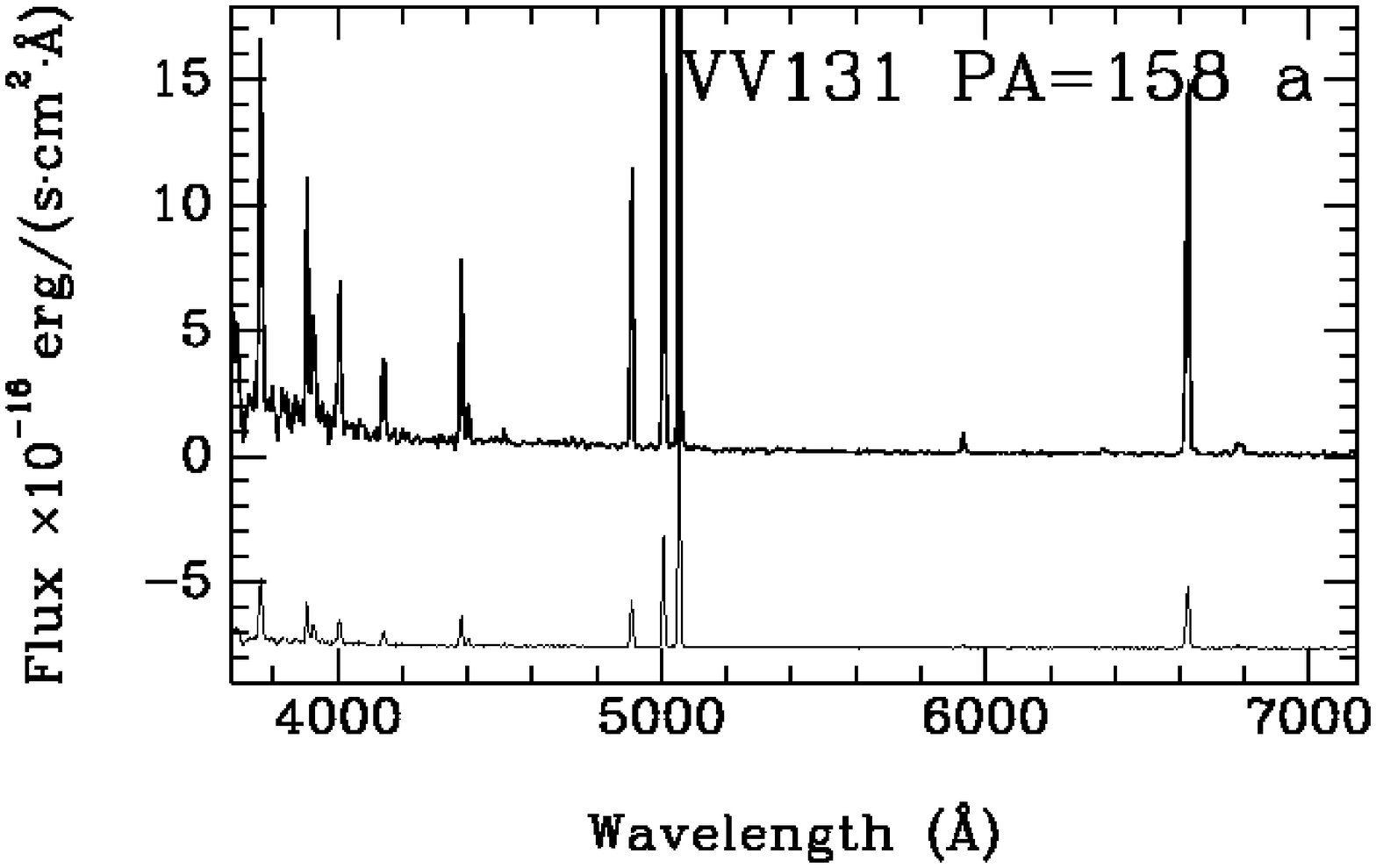}
   \caption[c]{{\it Top:} Blue DSS-2 image of VV~131 with the
	  positions of the long slit  superimposed.
          {\it Middle:} 2D spectrum of  VV~131 along $PA=158$\degr.
          {\it Bottom:} 1D spectrum of knot ``a'' in VV~131.
          }
     \label{VV131fig1}
   \end{figure}

   \begin{figure}
   \centering
   \includegraphics[angle=0,width=7.5cm,bb=28 135 560 670]{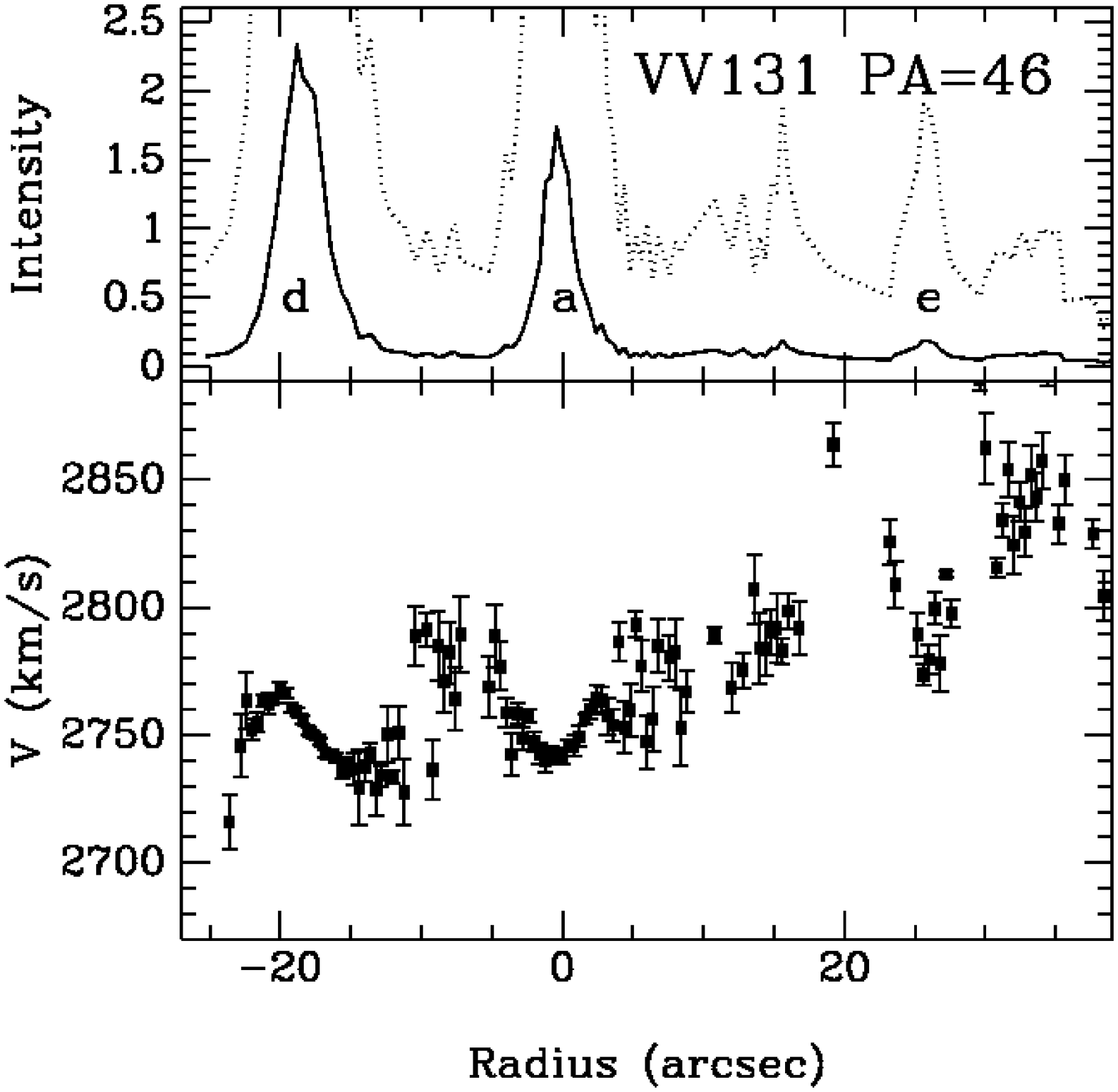}
   \includegraphics[angle=0,width=7.5cm,bb=28 135 560 670]{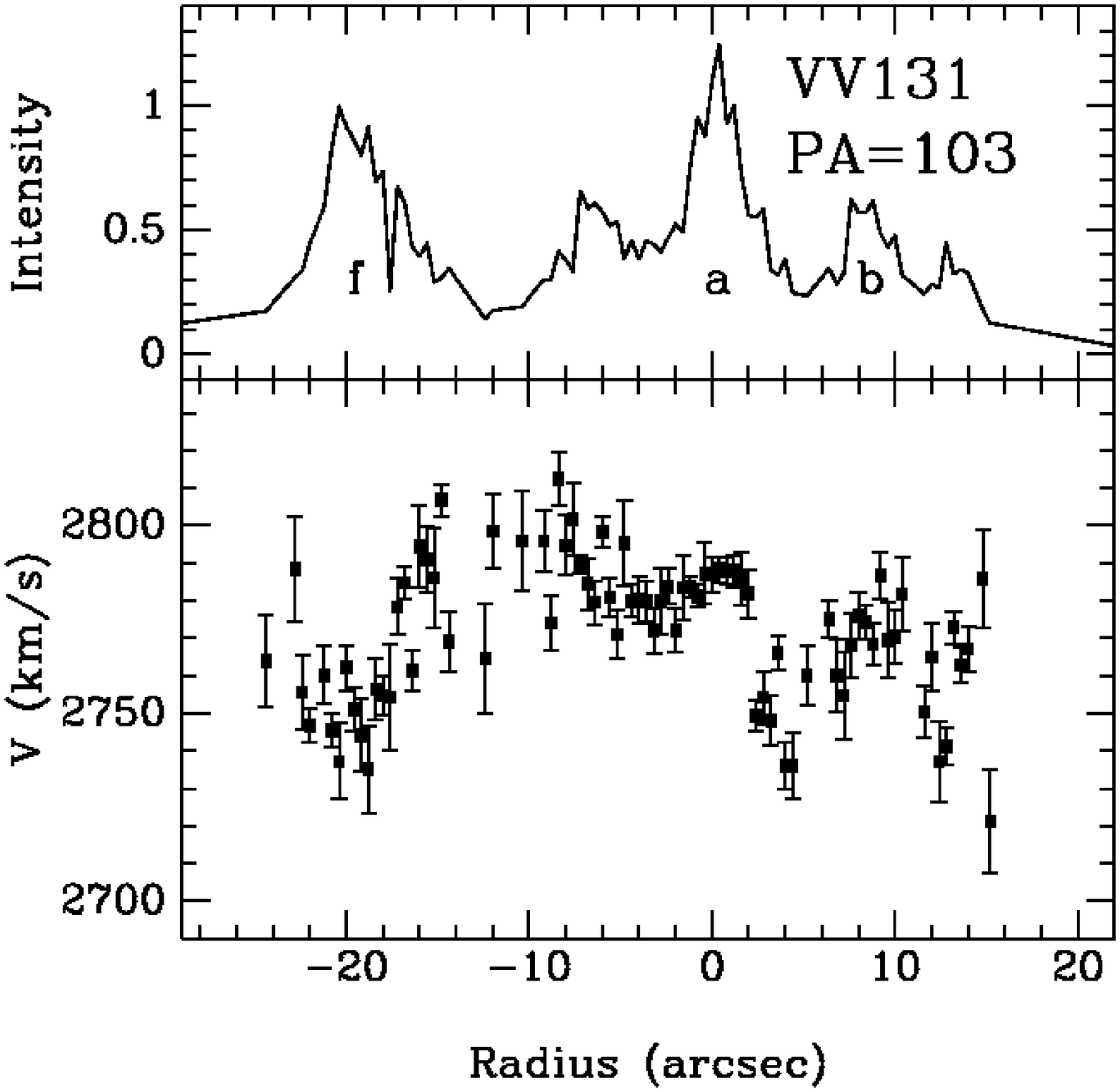}
   \includegraphics[angle=0,width=7.5cm,bb=28 105 560 670]{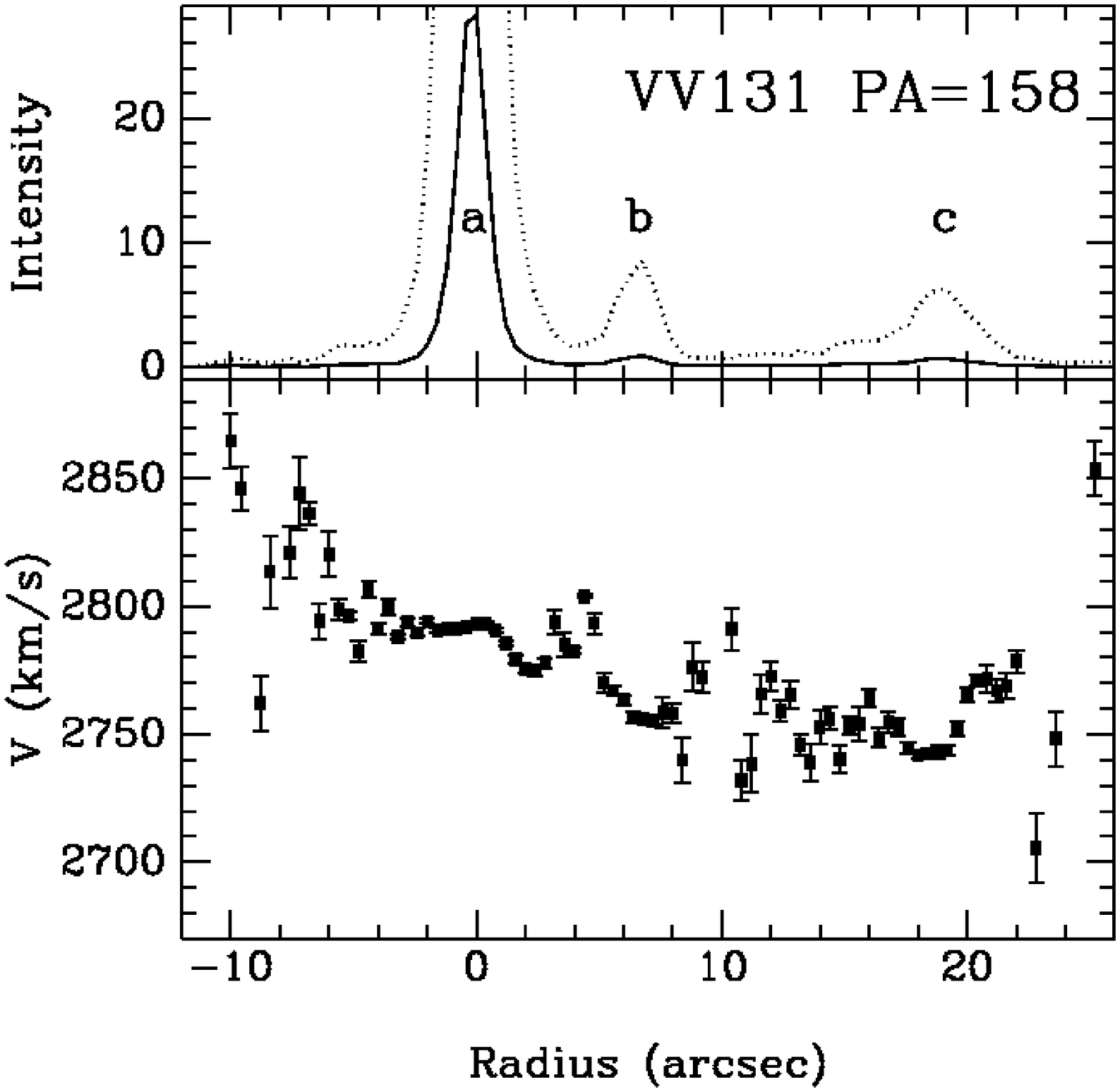}
   \caption[c]{
          Intensity profiles of H$\alpha$  for  VV~131
          along the slit with $PA$=46\degr, 103\degr\ and 158\degr, and
          respective P--V diagrams.
          }
     \label{VV131fig2}
   \end{figure}

   \begin{figure}
   \centering
   \includegraphics[angle=0,width=8.0cm,bb=20 105 580 690]{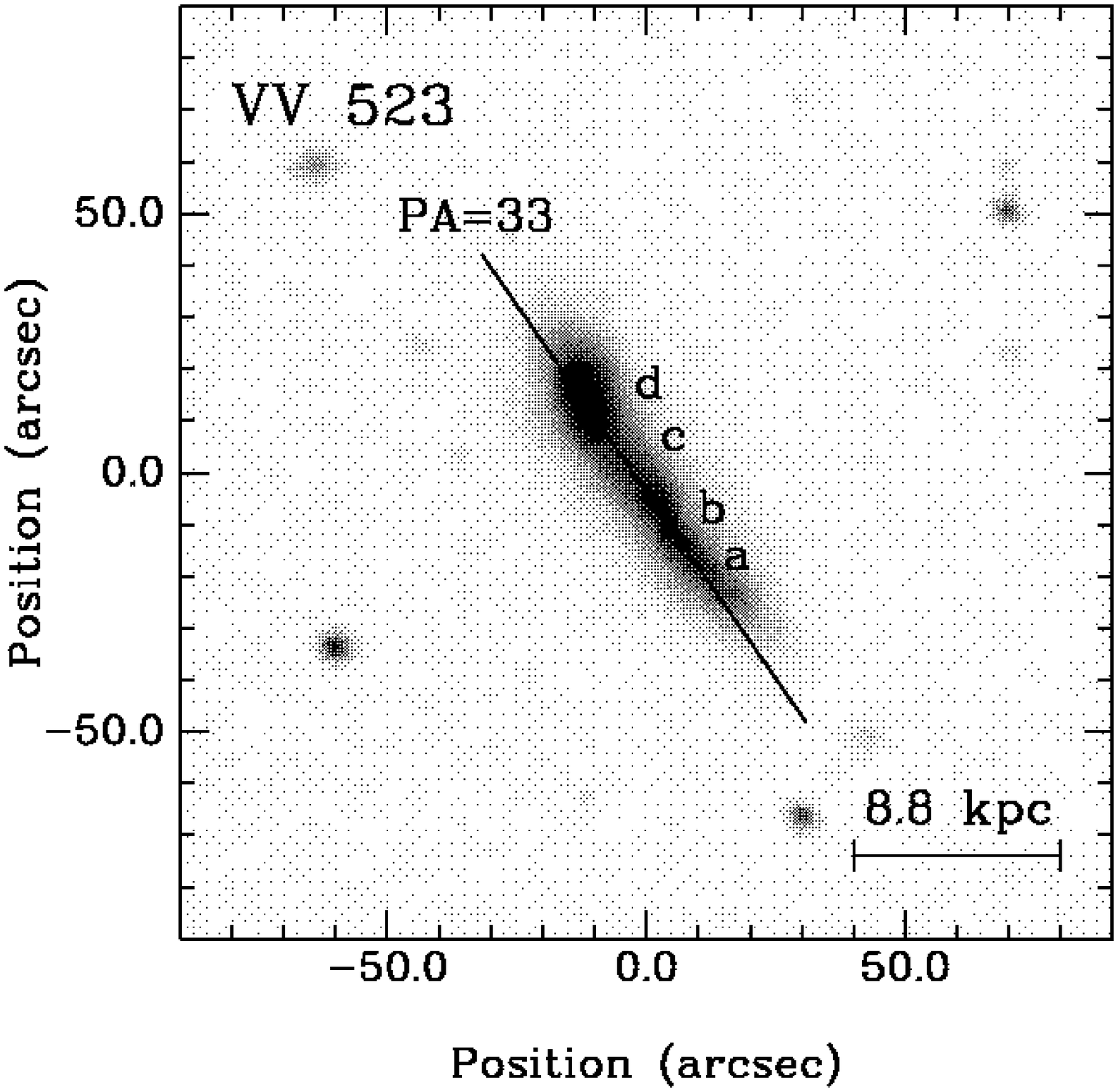}
   \includegraphics[angle=0,width=8.0cm,bb=20 170 580 590]{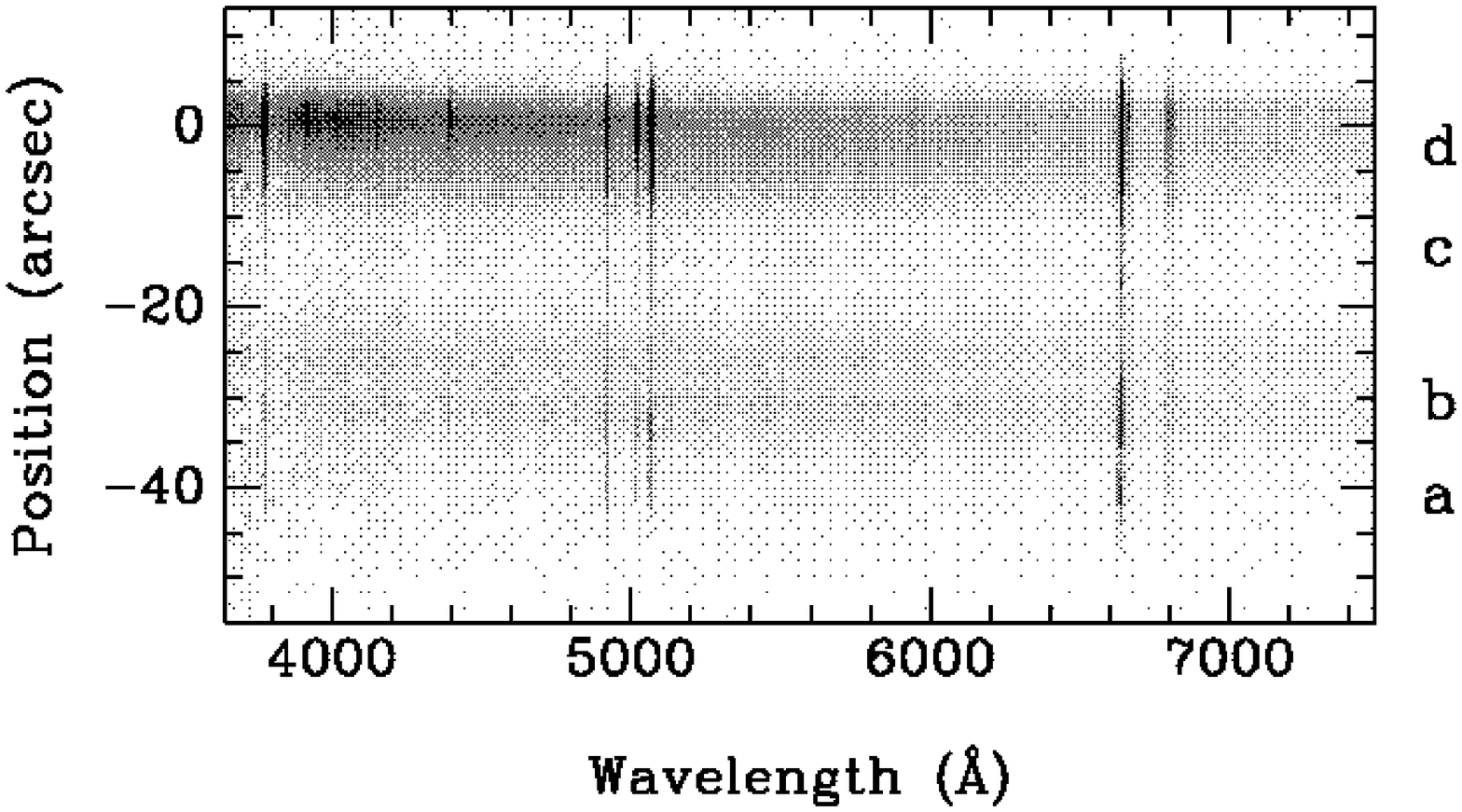}
   \includegraphics[angle=0,width=8.0cm,bb=20 165 580 540]{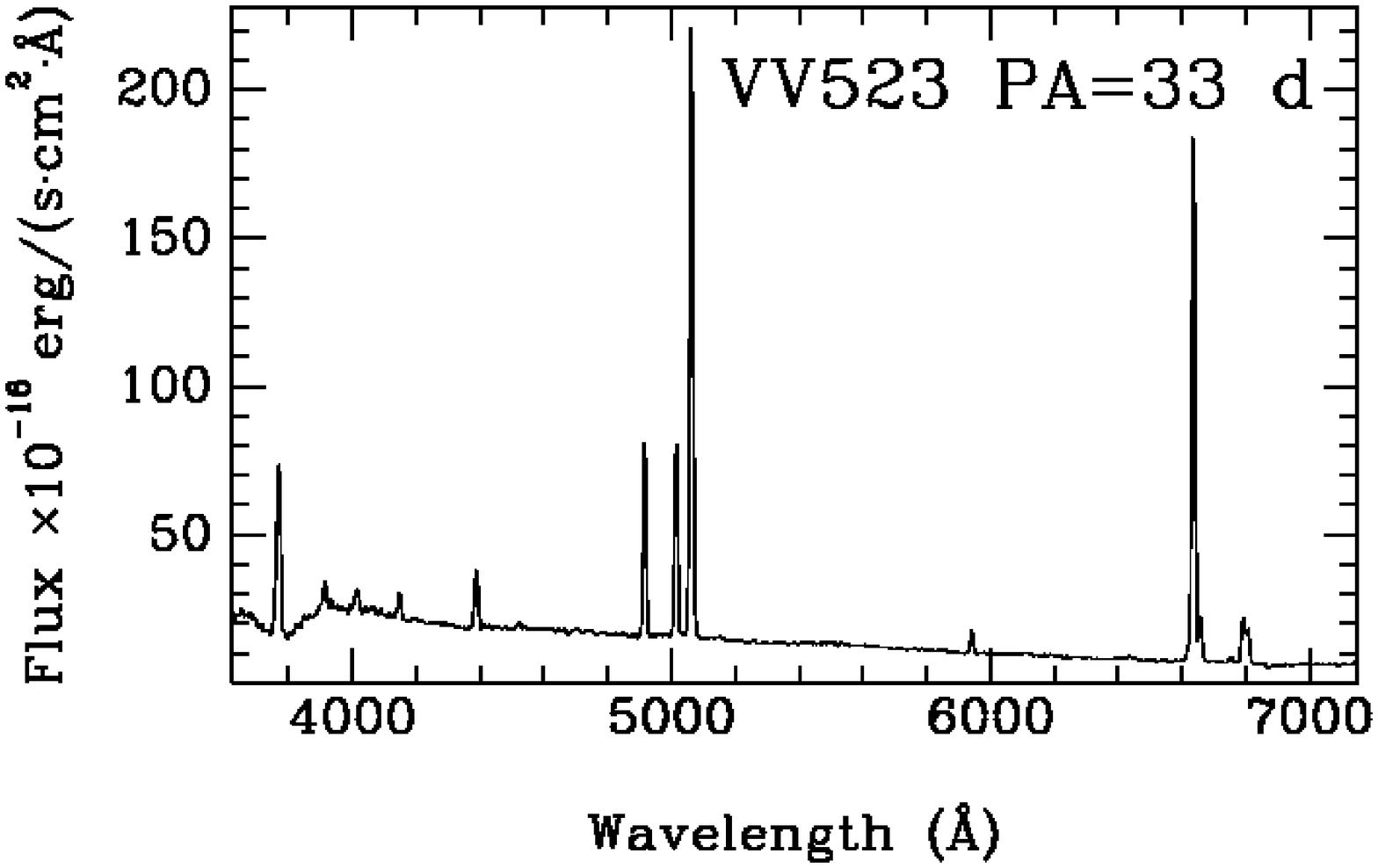}
   \caption[c]{{\it Top:} Blue DSS-2 image of galaxy VV~523 with the position
	   of the long slit  superimposed.
           {\it Middle:} 2D spectrum of VV~523 along $PA=33$\degr.
           {\it Bottom:} 1D spectrum of the brightest knot ``d''.}
     \label{VV523fig1}
   \end{figure}

   \begin{figure}
   \centering
   \includegraphics[angle=0,width=7.5cm,bb=28 105 560 670]{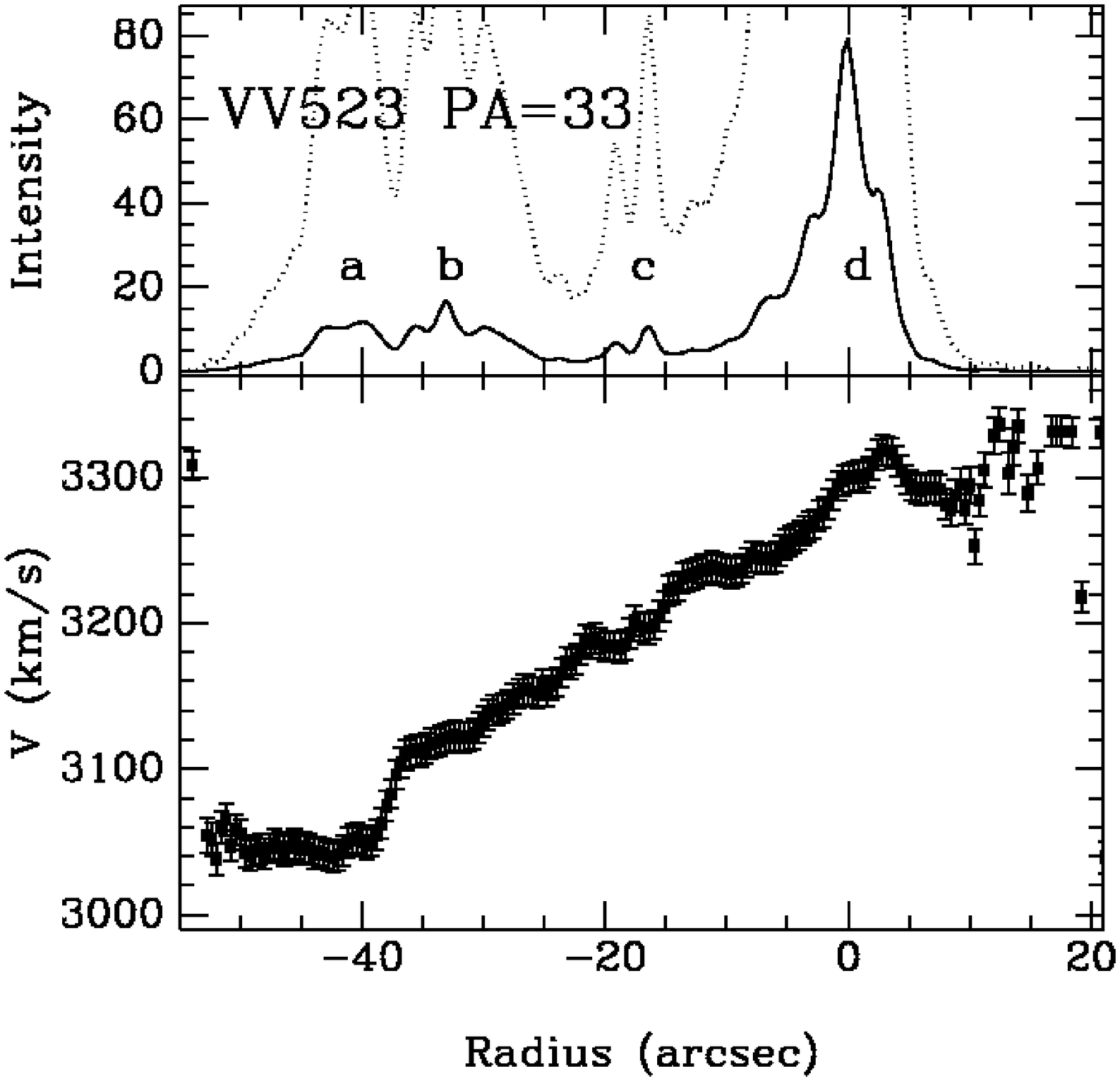}
   \caption[c]{
        Intensity profile of H$\alpha$ for VV~523
        along the slit with $PA=33$\degr, and respective P--V diagram.
          }
     \label{VV523fig2}
   \end{figure}

\subsection{VV~531 = NGC~1156 = UGC~2455}
\label{531_mult}

\subsubsection{General characteristics}

VV~531 is a rather bright  galaxy of IBm type with a strong
emission spectrum. Unlike the other galaxies considered in this paper,
VV~531 is classified as well isolated, being in the Karachentseva et
al. (\cite{Kara86})  list of isolated galaxies.  The distance to this galaxy,
estimated from its individual brightest stars, is 7.8 Mpc (Karachentsev et
al. \cite{Kara96}). These authors claim that this galaxy can be
considered as "the best example of an actually isolated galaxy in a general
field".

The size of the bright inner part of VV~531, containing numerous \ion{H}{ii}
regions, is  $\sim$60\arcsec\ (2.3  kpc).
Elmegreen \& Salzer (\cite{Elm99}), using H$\alpha$ observations,
marked about 20 large \ion{H}{ii} complexes in this galaxy. This
actively star-forming area is surrounded by an extended
envelope of low surface brightness, which makes the galaxy rather large:
its optical diameter $D_{\rm 25}$ exceeds 7 kpc.

Integrated blue colour indexes  of VV~531 obtained
by Esipov et al. (\cite{Esipov91}) ($(B-V)_\mathrm{0}$=0\fm56,
$(U-B)_\mathrm{0}$=-0\fm40) are consistent with a strong starburst in this
galaxy. In many aspects VV~531 is reminiscent VV~080. Their general
structure, luminosities, and axial ratios are rather similar.

A high S/N 21-cm line \ion{H}{i} profile and integrated
\ion{H}{i}-flux is presented by Haynes et al. (\cite{Haynes98}),
from which we derive its full \ion{H}{i} mass. A detailed \ion{H}{i}
map of this galaxy was obtained with the Westerbork Synthesizing
Radio Telescope (WSRT)  and some of the results were given by Broeils
\& van Woerden (\cite{BvW94}). For more details, they are
presented on the web-site of Westerbork \ion{H}{i} survey of
spiral and irregular galaxies
(WHISP).\footnote{http://www.astro.rug.nl/$\sim$whisp/Database/
  ~~~~~~~OverviewCatalog/ListByName/U2455/u2455plot.gif}
The description below is based on the figures from this web-page.
Its full extent on
the column density of 2$\times10^{19}$~atom~cm$^{-2}$ is
$\sim$9\arcmin$\times$7\arcmin. The major axis orientation of
\ion{H}{i} disk is close to the Eeast-West direction. The \ion{H}{i} velocity
field shows clear rotation with the velocity amplitude of
$\sim$50~\kms\ at the distance of $\sim$2\arcmin\ from the galaxy
center. Both data by Haynes et al. (\cite{Haynes98}) and from the
WHISP show a small but clear asymmetry in its global \ion{H}{i}
profile, appearing as some excess flux on the approaching side
of the profile at $V_{\rm rad} \sim$320--330~\kms. This
indicates either a faint additional source in the telescope beam,
or disturbed motion in the galaxy. Such a feature (at
$\sim$3\arcmin\ to the East of the galaxy center) is well
retraced on both \ion{H}{i} brightness and velocity maps with
full resolution (left column), and in the P--V diagram along major
axis (upper-right box).  We return to this point in
Section~\ref{VV531env}.

\subsubsection{Spectral properties}

Two spectra were obtained using both higher and lower spectral
resolutions for slit $PA$s of 42\degr\ and 96\degr\ (see Figure
\ref{VV531fig1}). The former orientation is close to the direction of the
VV~531 optical major axis. Relative line intensities
for its brightest knot `e' are presented in Table~\ref{Tab4}.
The H$\alpha$ over H$\beta$ intensity ratio reveals significant extinction:
C(H$\beta$) = 0.47$\pm$0.10 dex.  The latter corresponds to $E(B-V)$ =
0\fm32$\pm$0\fm07.
This value is rather consistent with the foreground Galaxy
extinction $E(B-V)$ = 0\fm22 in the direction of VV~531, as indicated in NED
according to Schlegel et al. (\cite{Schlegel98}). This in turn implies that
if additional extinction is present in that \ion{H}{ii} region, its
internal C(H$\beta$) at most amounts to $\sim0.15-0.20$ dex.

The estimate of the O/H  value shows that this galaxy is moderately
poor in heavy elements: we found for knot ``e'' by  the classical method
$12+\log$(O/H) = 8.17$\pm0.10$, which is close to the value 8.23 derived
for this galaxy by Vigroux et al. (\cite{Vigroux87}).

SF bursts across the body of VV~531 are rather well synchronized.
Ages of instantaneous starbursts (derived with the same IMF parameters, as
for VV~080 and VV~499 above)
in the four knots picked up in our observations, as estimated
from the $EW$(H$\beta$) (see Table~\ref{Tab5}), fall in the range of 4.0
to 7.3 Myr. The age of the youngest knot ``e'', situated near the galaxy
center, is consistent with the discovery of WR spectral
features in it (Ho et al. \cite{Ho95}). Unfortunately, the S/N in our
spectrum of this knot is not sufficient to confirm this result.

\subsubsection{Kinematics of gas}

Fig.~\ref{VV531fig2} shows the line-of-sight velocity
distributions along the $PA =$ 42\degr\ and 96\degr.  It is worth
noting  that the major axis orientations in
optical light ($PA =$ 42\degr) and in the \ion{H}{i} radio image
($PA\sim$90\degr) differ  significantly.   For both $PA$s,
global velocity gradients of ionized gas cannot be reliably
determined: they are masked by significant velocity swings
presumably due to the appearance of supershells near SF regions (see
section~\ref{VV131_kin} for a similar case in VV~131). The velocity
amplitude of such variations in VV~531 reaches $\sim$50~\kms.

The small velocity gradient agrees with the earlier estimate of the
maximal line-of-sight component of the velocity of rotation at radial
distances $\pm 1'$ (about 25~\kms)  obtained from H$\alpha$
observations by Karachentsev \& Petit (\cite{Kara90}). Hence,
the observed small-scale velocity irregularities of
ionized gas are comparable to or exceed the amplitude of the rotation
velocity in the inner region with a diameter of
1.5\arcmin\--2.0\arcmin. This is fully confirmed by the P--V diagram
along the major axis of the \ion{H}{i} WSRT image.

The total mass estimate within the optical diameter can be found from
the \ion{H}{i} linewidth W$_{\rm 20}$. Half of its value can be
accepted as an upper limit of circular rotation velocity for this
radius. With the values adopted from LEDA of
$W_{\rm 20}$=113~\kms\  and inclination  angle $i$=42\degr,
one may obtain $M_{\rm t}$ (within $D_{\rm 25}$)
$\approx$6$\cdot10^9~M$\sunn. Then the upper limits of
$M_{\rm t} / L_{\rm B}$ and $M_{\rm HI}$/$M_{\rm t}$ ratios are about 2
and 0.2, consistent with the range of these parameters for
Irr galaxies.  The results will not change much if we take
$W_{\rm 20}$=130~\kms\ from \ion{H}{i} profile obtained  at WSRT -
in agreement with their velocity field of this galaxy.

Note, however, that the density profile of \ion{H}{i} in VV~531 reveals
an inner, dense part (corresponding to the region of intense star
formation $R_{\rm sf}$$\approx$1\arcmin\ or 2.3~kpc) with a surface
density of (10--20)$M$\sunn~pc$^{-2}$  and a very extended rarefied gas
envelope exceeding the  optical size of the galaxy
(Broeils \& van Woerden \cite{BvW94}). If, following Karachentsev \& Petit
(\cite{Kara90}) and this work, we accept the value of 25~\kms\ as an upper
limit for the line-of-sight component of rotation velocity at
$R$$\sim$1\arcmin, the virial mass of this region is
$M(R_{\rm sf})$$<$8.5$\cdot10^8~M$\sunn. The gas surface density profile
shows that the hydrogen mass $M_{\rm HI}$ within this radius is
$\sim$(1.5--3)$\cdot10^8~M$\sunn, or (20--40)\% of the dynamical mass.
This implies
that active star formation in this galaxy is related to the region of very
high concentration of neutral gas. This is similar to the conclusions reached
by Taylor et al. (\cite{Taylor94}) and van Zee et al. (\cite{vanZee01}).

\subsubsection{Environment properties}
\label{VV531env}

Since VV~531, as mentioned above, is well isolated, its environment, by
definition, is not rich.
In particular, Karachentsev et al. (\cite{Kara96}) indicate it as the nearest
to VV~531 galaxies UGC~2684 and UGC~2716 with close radial velocities,
situated at projected distances more than 10\degr, which corresponds to
linear
distances of more than 1.3 Mpc. Such isolation, if real, would give one more
example of strong spontaneous SF activity in gas-rich dwarf galaxies and
form the basis for more active analysis  of non-interaction triggers of
starbursts.

Note, however, that visual isolation criteria based on catalogs
of known rather bright galaxies is in general not sufficient to address
this issue. Many instructive examples of detection of lower mass
companions with sufficiently strong tidal action near  ``isolated''
galaxies with enhanced SF (e.g., Taylor et al. \cite{Taylor93,Taylor95};
Pustilnik et al. \cite{Pustilnik01}) indicate that this question needs
more careful consideration. Even actively star-forming galaxies (BCGs),
situated in voids, delineated by bright galaxies, often have nearby low-mass
companions
disturbing their gas disks (e.g., Pustilnik et al. \cite{Pvoids02,LSB}).

In particular, some indications of possible low-mass companions of VV~531
appeared in the so-called ``Arecibo Slice'' -- a blind \ion{H}{i} survey by
Spitzak \& Schneider (\cite{Spitzak98}). This survey is a $\sim$1\degr\
wide declination strip centered at the declination $\sim$1.7\degr\ to the
South of the
position of VV~531. It detected 2 rather faint sources (designated in NED as
[SS98] 73 and LEDA 169969) at 1.5\degr\ and 3.2\degr\ from the position of
VV~531, with
radial velocities of 411 and 469~\kms, and estimated \ion{H}{i}-masses
of 1.0 and 1.7$\times10^{7}~M$\sunn\ (for accepted distance of 8~Mpc).
Their projected distances are much lower than those for the UGC galaxies
mentioned above, only $\sim$200 and 400 kpc.  Having in mind that the
``Arecibo Slice'' probed only a small part of the sky around VV~531, and
despite this detected two dwarfs in the  vicinity of VV~531, we
can expect that other similar dwarfs do exist in this region.
Such low-mass ``disturbers'',
in order to affect significantly more massive galaxies,
should either encounter them very closely (some ten kpc),
or experience the so-called sinking satellite merger.

As we already noticed above, there is clear evidence of a small elongated
\ion{H}{i} feature responsible for the \ion{H}{i} line asymmetry
(Haynes et al. \cite{Haynes98}; WHISP)
which may be naturally explained if a small and very nearby object
interferes with VV~531.
This object is at $\sim$7 kpc from the VV~531 center, and one would naturally
consider it as a sinking small satellite. Its gravitational
action on the gas disk of VV~531 should be rather strong
(e.g., Mihos \& Hernquist \cite{MH94}). So, current multiple starbursts are
probably being triggered by this object.

\subsection{VV~499 = DDO~053 = UGC~4459}
\label{499_mult}

\subsubsection{General characteristics}

VV~499 is a 14\fm5 Irr-galaxy  of unusually blue colour (according to LEDA,
its extinction-corrected integrated
$(U-B)$ colour is -0.48). Its integrated colour $(B-V)_{\rm tot}$=0.45,
according to Makarova (\cite{Makarova99},\cite{Makarova00}), is, however,
rather typical of dI galaxies.
This galaxy belongs to the Low Surface Brightness (LSB) class, since its
central brightness $\mu^{0}_{\rm B}$=24.1 mag. arcsec$^{-2}$ (Makarova
\cite{Makarova99},\cite{Makarova00}).

This is a dwarf galaxy of very low luminosity ($M_{\rm B}$
\mbox{$\approx$--13\fm6})
and very small optical diameter ($\sim$1 kpc). Emission regions of VV~499
are situated along two parallel diffuse strips, with the western one of
significantly lower surface brightness. Red  DSS-2
image shows extended  low brightness background between the
strips and around them. Blue stars are observed not only in
the strips, where their concentration is higher, but also in between
(Schulte-Ladbeck \& Hopp \cite{Regina98}).
The highly irregular structure of star-forming regions is clearly
seen in the HST image (see HST archive).

\subsubsection{Spectrophotometry}

Several spectra were obtained with the
slit oriented at PA = 98\degr, 140\degr\ and 145\degr\ (see Figures
\ref{VV499fig1} and \ref{VV499fig2}).  Line intensity ratios
in knot ``b'' correspond to high excitation gas ($T_{\rm e} \sim$19000~K).
Bright emission lines detected in several knots over the galaxy body
indicate that the galaxy has experienced recent starburst. According to
Leitherer et al. (\cite{Starburst99}), the observed EWs(H$\beta$) in knots
``a'',  ``b'' and ``d'' (see Tab.~\ref{Tab5}) correspond to the ages of
instantaneous
starbursts (with Salpeter IMF and $M_{\rm low}$ and $M_{\rm up}$ of
1~$M$\sunn\ and 100~$M$\sunn, respectively) of $\sim$3--8 Myr. For
$M_{\rm low}$=0.1~$M$\sunn\ the observed EWs(H$\beta$) will correspond to
even smaller ages.
For the brightest knot ``b'' we derive value of  O/H with the T$_{\rm e}$
method -- 12+$\log$(O/H)= 7.52$\pm$0.08, that is consistent with the
value of 7.62 obtained by Skillman et al. (\cite{Skillman89})
by the empirical method.
Notice that luminosity and O/H of DDO~053  match the ``O/H versus
$L_{\rm B}$'' relationship for dIrr galaxies suggested by  Skillman et al.
(\cite{Skillman89}).

The oxygen abundance of VV~499 assigns it to the
group of very rare extremely metal-deficient gas-rich dwarfs.
Izotov \& Thuan (\cite{Izotov99}) argued that among such
galaxies there may exist truly young local galaxies experiencing
the first episode of star formation. However, the
analysis of resolved stellar population from HST data indicates the
presence of old RGB stars (Karachentsev et al. \cite{Kara02}), which implies
that VV~499 is at least 3-Gyr old
(L. Makarova, private communication). Its very low metallicity is probably
a result of significant metal loss due to metal-enriched galactic wind.

\subsubsection{Gas kinematics }

Despite the observed appreciable small-scale velocity variations
(Figure~\ref{VV499fig2}),
a significant velocity gradient can be traced along the slit at
$PA$=140\degr\--145\degr\ (nearly parallel to the major axis). The observed
amplitude amounts to 40~\kms\ for full extent of 65\arcsec.
This is consistent with the observed FWHM of \ion{H}{i} profile
$W_{\rm 50}$=30~\kms\ (Huchtmeier \& Richter \cite{Huchtmeier89}).
Taking into account that the slit did not pass through the center of the
outer isophotes,  and also the possible role of velocity dispersion,
one can consider this amplitude as the lower limit of the
expected line-of-sight  component of circular velocity gradient. After
the inclination correction (according to LEDA $i$$\sim$45\degr), we obtain
the amplitude of circular  rotation of $V_{\rm m}$$\sim$28~\kms.
It results in the total mass within the radius 0.65 kpc
$M_{\rm t}$$\sim$1.2$\cdot$10$^{8}~M$\sunn\
and $M_{\rm t}$/$L_{\rm B}$$\sim$3 solar units.

The galaxy is quite a gas-rich object.
The total mass of \ion{H}{i} (see Table~\ref{Tab1}) is
7.1$\cdot$10$^{7}~M$\sunn.
Accounting for the mass fraction of \ion{He}{}, we conclude that the gas
mass is close to  dynamical mass inside the optical diameter.

\subsubsection{Environment and interactions}

VV~499 is  a member of the M~81 galaxy group, although there are no
neighbouring  galaxies of comparable luminosity within several
optical diameters around it.
The most recent HST Colour-Magnitude diagram-based (CMD) estimate of its
distance is 3.56~Mpc (Karachentsev et al. \cite{Kara02}).

The M~81 group consists of two subgroups: M~81 itself at 3.7 Mpc from the
Local Group (LG), with a radial velocity relative to the LG of
$V_{\rm LG}$=108~\kms, and a subgroup of NGC~2403, at 3.2 Mpc and
$V_{\rm LG}$=267~\kms, with the
distance between these subgroups $\sim$1.0~Mpc (Karachentsev et al.
\cite{Kara02}).
VV~499, with its $V_{\rm LG}$=150~\kms\ and $D$=3.56~Mpc seems to belong
to the M~81 subgroup. It is situated on the periphery of this subgroup, about
8.4\degr\ to the
South-West of M~81, or $\sim$0.5 Mpc in projection, in the direction of
subgroup of NGC~2403. Its distances to both subgroups are comparable.
Tidal effect onto VV~499 of any separate galaxy belonging to these
subgroups  could be rather subtle. However, the tidal role of the whole
subgroup M~81 (with a total mass of $\sim$10$^{12}~M$\sunn, Karachentsev et
al.~\cite{Kara02}) is quite important (see, e.g., estimates in Pustilnik et
al.~\cite{Pustilnik01}). The case of
enhanced star formation in VV~499 probably has a relation to statistics
of distant dwarf satellites of massive spirals, studied by
Zaritsky et al.
(\cite{Zaritsky97}). Most of those satellites are emission-line galaxies.
We suggest that the likely trigger of current starburst in VV~499 is its
tidal disturbance by M~81 subgroup(s) or/and interaction with
intragroup gas.

\subsection{VV~131 = UGC~4874 = PGC~26104}

\subsubsection{General characteristics}

VV~131 is a rather small and poorly studied galaxy of SBm type. Its total
magnitude is quite uncertain. The LEDA database gives
$B_{\rm T}$=17\fm6$\pm$0\fm2, or corrected
$B_{\rm c}$=16\fm8, although there is some  evidence that these values
may be
related only to the inner, brightest part of the galaxy.  VV and UGC catalogs
give uncorrected $m_{\rm b}$=16\fm0. This is close to
$B$=16\fm1, derived from the APM $B$-magnitude, applying the empirical
relation
between calibrated CCD $B$-magnitudes and APM $B$-magnitudes, obtained by
Kniazev et al. (\cite{Kniazev02}). Below we use
$B_{\rm tot}$=16\fm0$\pm$0\fm5  (see Table~\ref{Tab1}).

Visually the galaxy looks like as inclined letter ``$\Theta$'' with a short
bright bar and barely visible outer ellipse of low
brightness with the major axis diameter of $\sim$1\arcmin.

There are several bright knots along the ring, one of
which, located  at the NW edge (knot ``c'' in
Fig.~\ref{VV131fig1}) has its own PGC number (2790837).
Nevertheless, it is quite evident that knot ``c'' is
just one of several bright knots of the same galaxy.
From our three slit positions we also have not found evidence of a
probable star superimposed on this galaxy, as suggested in the
UGC. At the adopted distance (39.6 Mpc), the linear diameter of the
ring is $\sim$11 kpc. It seems that this ring was ignored while
determining the position of the major axis and isophotal diameter
$D_{\rm 25}$ of the galaxy, presented in the LEDA database. If we
use the parameters presented in Table~\ref{Tab1}, the total corrected
absolute $B$-magnitude of the galaxy  is $M_{\rm B}$=--17\fm1.
Hence,  VV~131 can be considered as a non-typical dwarf  (by its
luminosity) galaxy, possessing unusually regular structural
features: a symmetric bar and a low surface brightness ring.
Curiously, it looks very similar to the other 15-magnitude SBm
galaxy UGC 634 (its nice $B$-band image is given by van Zee
\cite{vanZee00}). By analogy, VV~131 may also be classified as a
dwarf low surface brightness barred spiral galaxy.

According to Schneider et al. (\cite{Schneider92}), the \ion{H}{i}
integrated flux of VV~131 is 9.3 Jy$\cdot$\kms, which corresponds
to total mass  $M_{\rm HI}$=3.45$\cdot$10$^9~M$\sunn.
The $M_{\rm HI}$/$L_{\rm B}$ ratio for this galaxy is unusually high
($\sim$3 solar units), exceeding characteristic values for normal
late-type galaxies by about a factor of six. Partly it may reflect
the underestimation of the galaxy light in its low surface
brightness extended outer regions.  Note, however, that the
formally found mean surface density of \ion{H}{i} of VV 131
within its optical diameter (11 kpc) is unrealistically high:
 4$M_{\rm HI}$/$\pi D_{\rm opt}^2$$\approx$36~$M$\sunn~pc$^{-2}$,
which is more typical for the
densest inner parts of blue compact dwarfs (van Zee et al.
\cite{vanZee01}). Is this galaxy surrounded by an extended
\ion{H}{i} cloud? The  overestimation of \ion{H}{i} mass of
VV~131 due to the  probable confusion from the nearby galaxy
UGC~4868 seems not to play a significant role. Both galaxies
were clearly detected in the \ion{H}{i} 21~cm line independently,  and
the measured integrated \ion{H}{i} flux of UGC~4868 is about a
half of that for VV~131 (Schneider et al. \cite{Schneider91}).
Thus, the unusually high gas content of this galaxy deserves
further detailed study.

\subsubsection{Element abundances}

The presence of the faint [\ion{O}{iii}] line 4363~\AA\
(see Figure~\ref{VV131fig1}) allows us directly to estimate the electron
temperature
$T_{\rm e}$ of \ion{H}{ii} regions and to determine O/H. It was
found to be 12+$\log$(O/H)=7.85$\pm$0.06 ($Z$$\sim$1/12~$Z$\sunn)\footnote{We
adopt solar oxygen abundance as
12+$\log$(O/H)=8.92, according to Anders \& Grevesse (\cite{Anders89}).}.
Hence, VV~131 may be considered  as  a low-metallicity dwarf ring galaxy.
The element ratios Ne/O and N/O (Table~\ref{Tab3}) for this region are
consistent with the typical values for metal-poor galaxies, --0.7 and
--1.5 dex, respectively, as summarized by Izotov \& Thuan
(\cite{Izotov99}).

\subsubsection{Kinematics and mass estimation}
\label{VV131_kin}

We got for this galaxy P--V diagrams (Fig.~\ref{VV131fig2}) along the
slits with $PA=$46\degr, 103\degr\ and 158\degr\ (see Fig.~\ref{VV131fig1}).
All three slit positions went through the central part of the galaxy
(knot ``a'').  The other bright features are marked as ``b'', ``c'', ``d'',
``e'' and
``f'' in Fig.~\ref{VV131fig1}; the respective peaks in distribution
of H$\alpha$ intensity are marked in Fig.~\ref{VV131fig2}.
For knot ``a'', as expected, we derive a radial velocity of $2785\pm5$~\kms,
very close to the systemic  velocity of VV~131 presented in Table~\ref{Tab1}.

P--V diagrams for all three $PA$s show significant deviations from
smooth curves, which correlate in positions with peaks in distribution
of H$\alpha$ intensity. Characteristic sizes (FWZI) of such deviations  are
$\sim$5\arcsec--10\arcsec, corresponding to linear sizes of 1--2~kpc.
Their velocity amplitudes are in the range of 20--40~\kms. These values
correspond to parameters of ionized gas supershells observed in many
galaxies with active SF (e.g., Marlowe et al.~\cite{Marlowe95}, Martin
\cite{Martin96},\cite{Martin97}).
Such supershells originate as a result of the cumulative action of massive
star winds and SNe related to
individual starbursts with a sufficiently large number of massive stars
(e.g. Pustilnik et al. \cite{P01_Granada},\cite{LSB}).

Global distribution of velocity along the galaxy body is best seen on the
P--V diagram along $PA=$46\degr, which is close to the direction along the
galaxy major axis. Despite the significant deviations near bright \ion{H}{ii}
regions, a gradient with full amplitude of $\sim$110--120~\kms\ over
$\sim$60\arcsec\ ($\sim$11.5~kpc) is clearly seen. It agrees remarkably well
with the full width on the level of the 20\% of maximum of the \ion{H}{i}
21-cm
line profile: $W_{\rm 20}$=114$\pm$7~\kms\ (Schneider et
al. \cite{Schneider91}).
Rotation curves, determined from mapping of dwarf galaxies in the 21-cm
line, are usually traced to much larger distances than the optical
radius of a galaxy. The coincidence of $W_{\rm 20}$ and H$\alpha$-velocity
amplitude implies that rotation curve of VV~131 reaches its maximal value
at the radii, close to those sampled by our long-slit spectrum.

Adopting that the internal form of observed ring is a circle, from the
apparent
ratio $b/a$=0.5 we can determine inclination angle $i$$\sim$60\degr. Then,
the true rotation velocity will read as a half-amplitude (57~\kms) divided
by $\sin i$, that is $V_{\rm rot} \sim66$~\kms.

The total mass $M_{\rm t}$ of a galaxy, defined as
$V_{\rm rot}^2$$\cdot$$R/G$,  within the radius of 6 kpc is
(6.0$\pm$1.2)$\cdot$10$^9~M$\sunn.               
This estimate implies that neutral gas (whose total mass,
including the He mass-fraction of 0.25, is 4.6$\cdot$10$^9~M$\sunn)
can be one of the main dynamical components inside the radius of 6 kpc.
However, while the radial density distribution of neutral gas is
unknown, it is possible that only a fraction of gas mass is
bounded in this region.

Thus, we may conclude that VV~131 is a low-luminosity low-mass
galaxy with bar and ring structures, with SF activity taking
place both near the center (along the bar) and in the ring.

\subsubsection{Environment and interaction}

The abovementioned neighbouring galaxy UGC~4868 is classified also as SBm
(e.g., Schneider et al. \cite{Schneider91}). This galaxy has  about the same
luminosity and very similar width of \ion{H}{i} profile
($W_{\rm 20}$=120~\kms\ versus 114~\kms\ for VV~131). Hence, the total
masses of these two galaxies are comparable.
Its location at the projected distance of only $\sim$6\arcmin\
(65 kpc, or 6 optical diameters) to the SW of VV~131, and the small relative
radial velocity ($\Delta V$=1$\pm$7~\kms) suggests that VV~131 was disturbed
by this galaxy during their recent encounter. Bar structures can quite easy
form in HSB (High Surface Brightness) disks as a result of strong
perturbations during close encounter (e.g., Mihos et al. \cite{Mihos97}).
Subsequently, bars can trigger the enhanced SF activity in these galaxies.
The observed bar in VV~131 and several knots with significant SF
seem to well match model expectations.

For UGC~4868,  also having significant
a bar and highly disturbed outermost isophotes, no prominent SF
bursts are seen. Probably this is related to different delay time for
development of strong perturbation and subsequent gas collapse after the
moment of pericenter passage in these two dwarf galaxies.
The latter is related to rotation periods of the galaxies involved
(T$\sim$600~Myr), and can easily differ by several tens to a hundred Myr.
As regarding unusual structures, seen in dwarf galaxy VV~131, we suggest
that it is a transient structure appearing as a result of recent strong
disturbance.

\subsection{VV~523 = NGC~3991 = UGC~6933}
\label{523_mult}

\subsubsection{General characteristics}

VV~523  belongs to the category of clumpy irregulars (Casini \&
Heidmann \cite{CH76}), which are characterized by very clumpy
structure and by active star formation. Together with the nearby
NGC~3994 and NGC~3995 it forms a triplet of interacting galaxies.
The galaxy is seen nearly edge-on, and unlike most disky
galaxies, it does not have a noticeable bulge. By this reason
it looks like a chain of small galaxies in contact. Its morphology
also resembles some comet-like blue compact galaxies. Spectral
observations  clearly indicate that this is a single object.
A strong high excitation emission-line spectrum and unusually blue
colour of VV~523: ($U-B$=--0\fm33), along with high FIR
luminosity, far exceeding the optical one (Hecquet et
al.~\cite{Hecquet95}), reveal very intense star formation all
over the galaxy. Its blue luminosity is unusually high for Irr
galaxies ($M_{\rm B}$$\approx$--21\fm0), especially if we account
for some internal extinction due to its almost edge-on orientation.
Evidently its outstanding blue luminosity is related to its young
(5--6 Myr, see Table~\ref{Tab5}) powerful starburst.

Optical observations of this galaxy were described in several papers. The
most detailed investigations were presented by Hecquet et al.
(\cite{Hecquet95}, see also references there to earlier works).
These authors carried out $BRI$-photometry (including
photometry of individual clumps) and long-slit spectroscopy to
study kinematics, physical conditions in ionized gas and its oxygen
abundance.
Using evolutionary population synthesis models, they came to the
conclusion that evolution of this galaxy is characterized by
discrete starbursts rather than by continuous star formation.

For \ion{H}{i} flux
we accept the mean of three measurements, obtained at Arecibo
 (17~Jy$\cdot$\kms) (Haynes et al.~\cite{Haynes98}).
Then  resulting $M_{\rm HI}$/$L_{\rm B}$=0.23 (Table~\ref{Tab1}) is
rather typical of irregular galaxies.

\subsubsection{Spectral observations}
\label{VV523_abun}

Two observations of VV~523 were carried out with the lower and higher
spectral resolutions. In both cases the slit was oriented along
the main body of the galaxy (see Figure~\ref{VV523fig1}).
Line ratios and other parameters of ionized gas are presented in Table
\ref{Tab4} for the brightest condensation ``d''.  Results of our
spectrophotometry are consistent in general with those from Hecquet et al.'s
(\cite{Hecquet95}) for this knot, but are more deviant for the fainter
components.
We concentrate further on analysis of the data for knot ``d'', where the
highest S/N ratio was reached. The largest difference between our and Hecquet
et al. spectrophotometry for this knot is for the intensity of
[\ion{O}{ii}] $\lambda$3727 line, which is a factor of 3 lower in our
spectrum in comparison to their value. These authors indicate that their
accuracy in blue reaches 30\%. We do not exclude that such large difference in
the [\ion{O}{ii}] $\lambda$3727 relative intensity is partly caused by
slightly different position angle of the long slit (our PA=33\degr\ and
their PA=31\degr)
and probably different slit width (not indicated in their paper).
However, since for the other knots we have similar large differences in the
relative intensities of this line, we suggest that their real errors of flux
calibration in UV were significantly higher.


Our estimates of O/H are quite uncertain, since the line
[\ion{O}{iii}] $\lambda$4363 in our spectrum is barely detected.
Hecquet et al. (\cite{Hecquet95}) give for knot ``d'' (their region ``C''),
based on the empirical method of Edmunds and Pagel (\cite{EP84}), the value
of 12+$\log$(O/H)=8.43$\pm$0.20.
Our estimate of this value by the classic method is of very low accuracy:
8.33$\pm$0.30. We used also the Pilyugin (\cite{Pilyugin01}) empirical
$P$-method, devised for ``high'' metallicity \ion{H}{ii} regions. With the use
of this method, 12+$\log$(O/H)=8.76. From these two values we combine some
`average' 12+$\log$(O/H)=8.65 with probable uncertainty of 0.2 dex.

Parameter $C(H\beta)$ in all four knots according to our measurements
is consistent within the uncertainties of the order of 0.1 with
the constant value of 0.10 dex. This implies (despite the edge-on orientation)
only a small attenuation of starburst light.
The data show that VV~523 fits in general to the $Z$ - $L_{\rm B}$
relation for Irr galaxies. Correction of its $L_{\rm B}$ for
significant enhancement due to current starburst will produce an even better
match.

\subsubsection{Gas kinematics}

The line-of-sight velocity distribution along the slit is shown in
Fig.~\ref{VV523fig2}.  Rotation velocity  changes linearly,
although some distortions of the P--V diagram with the amplitudes of
$\sim$30--40~\kms\ are noticeable in the regions of knots ``a''
and ``d'', suggesting the existence of ionized gas shells.

The observed velocity distribution along the whole body of VV~523
mimics a solid-body rotation, which is often observed in edge-on
late-type galaxies. This linearity may be naturally explained as
a result of inner absorption, which hides the \ion{H}{ii} regions
located near the line of nodes, where the Doppler component of
velocity of rotation is maximal. This absorption effect,
however,  should be small in the outer regions, so that the
velocity difference between the opposite edges of the galaxy
remains close to twice the  maximal velocity of
rotation.  Note also, that, as shown, e.g., by Keel
(\cite{Keel93}),  the large extent of the rising part of a galaxy
rotation curve  can be indicative of strong disturbance, which
changes its normal shape (usually it looks like a solid-body only
in the central part, and further from the center the velocity
gradient becomes small).

According to our data (see Fig.~\ref{VV523fig2}),   the full
amplitude of the velocity change along the slit on the extent of
73\arcsec\ is about 290~\kms. This is a bit higher than the value
of $W_{\rm 20}$=245$\pm$12~\kms\ from LEDA. Taking half of the
optical value given above as the rotation velocity of the outer
parts of VV~523, one can derive the total mass within its optical
diameter $D_{\rm 25}$=86\arcsec\ ($\sim$19 kpc). Note that the RC3
catalog gives $D_{\rm 25}$=87\arcsec, consistent with the
value in Table~\ref{Tab1}. This mass is
$M_{\rm t}$$\approx$4.6$\cdot$10$^{10}~M$\sunn. Since the appearance of
this galaxy is quite unusual, it is interesting to look for the ratio
$M_{\rm t}$/$L_{\rm B}$ in
comparison to other disk galaxies. Taken at face value it
amounts about 1.3. Presumably, some corrections should be applied
to account for the significant contribution of blue light from young
SF regions, on the one hand, and possible obscuration within the
galaxy disk, on the other hand. The visual appearance of the
brightest SF regions with respect to the rest of the galaxy body
implies that they contribute more than a half of the $B$-band total
light. This means that for the underlying stellar population
$M_{\rm t}$/$L_{\rm B}$ should be larger than $\sim$2.6.
On the other hand, due to its almost edge-on orientation, the
extinction in this underlying disk can reach $A_{B} \sim$1\fm0.
This implies that non-bursting $M_{\rm t}$/$L_{\rm B}$$\sim$1. This value is
compatible with the model values for stellar population of typical
late-type galaxies of near-solar metallicity and almost constant
SFR (e.g., Bell \& de Jong \cite{BdJ00}).

The hydrogen content appears to be rather high in VV~523. If all \ion{H}{i}
was concentrated within the optical radius, the ratio $M_{\rm HI}$/$M_{\rm t}$
would be $\sim$0.18.
Accounting for the helium mass fraction, the neutral gas mass fraction can
reach $\sim$0.24.
In reality its significant portion  can spread far beyond the optical radius.

\subsubsection{Environment and interactions}

As indicated above, VV~523 is a member of a triplet. Other members have
comparable luminosity and mass. The projection distance of VV~523
to its neighbours is $\sim$50~kpc.

Interaction-induced SF in sufficiently close systems,
including gas-rich disk galaxies, was studied in detail by several authors
(e.g., Keel \cite{Keel93}, Bernloehr \cite{Bernloehr93}).
In particular, for VV~523,  kinematics of ionized gas was analyzed
by Keel (\cite{Keel85},  \cite{Keel93}, \cite{Keel96}) among many other
galaxy pairs.
Compelling evidence of interaction-induced disturbances in VV~523
may include the abovementioned solid-body rotation curve for at least
a half of the visible disk extent and also the clearly seen optical disk
warps.
Hecquet et al. (\cite{Hecquet95}) come to similar conclusion on the
interaction-induced SF in this galaxy.

\section{Summary and conclusions}
\label{Conclusions}

One of the most important questions concerning the nature of  ``nest'' and
``chains'' type VV galaxies
is  their evolution status.  Current study, as well as several previous
publications, demonstrate that most of the low-luminosity VV galaxies
which appeared to be single, are relatively nearby irregular galaxies
with several bright knots of enhanced star formation.
As the line-of-sight velocity distributions along the slit reveal, such
knots are often accompanied
by gas outflows with velocities of several tens \kms\ and typical
sizes between 0.1 and 2.0 kpc.

Visual inspection of the images of these galaxies shows that unlike more
common irregular galaxies, in many single VV galaxies \ion{H}{ii} regions
are observed superimposed on the low surface brightness (LSB) background of
underlying disks. This may give visual impression of a multiple
system. The LSB nature of disks in some of these galaxies implies that the SF
rate averaged over the lifetime of these galaxies is much lower than the
present day one.

Several lines of evidence for the importance of weak interactions to trigger
star formation in gas-rich disk galaxies are suggested during the last one and
a half decade. Icke (\cite{Icke85}) probably was the first to have drawn
attention on the importance
of relatively weak interactions to trigger gravitational instability in gas
disks via generation of shocks. Much observational evidence for the
important role of weak interactions in triggering SF was obtained since that
time, including
detection of low mass \ion{H}{i} companions of nearby \ion{H}{ii}-galaxies
(e.g., Chengalur et al. {\cite{Chengalur95}; Taylor et al. \cite{Taylor93},
 \cite{Taylor95}; Taylor \cite{Taylor97}) and optical faint companions of BCGs
(Pustilnik et al. {\cite{Pustilnik01}, Noeske et al.
\cite{Noeske01}). Recent results on late spirals by Reshetnikov
\& Combes (\cite{Reshet97}) and Rudnick et al. (\cite{Rudnick00})
also suggest the importance of weak interactions in modulation of
the SF history of gas-rich galaxies.

Our analysis of several single
VV galaxies shows that weak interactions can be an important
factor shaping the appearance of this type of object.
In particular, such interactions should presumably trigger
starbursts in the galaxies VV~080 and VV~523, belonging to
groups. Close neighbour of comparable mass to the galaxy VV~131
may also be an efficient disturber to ignite active star
formation in the latter object. Finally, provisional
identification of a sinking satellite near VV~531 as a tidally
disturbing object can explain its synchronized SF in many
regions. All these examples indicate the importance of interactions
to trigger starbursts in the majority of VV galaxies studied.

From the observational data presented above and their
analysis, along with published data we draw some preliminary
conclusions:

\begin{itemize}

\item
Spectrophotometry and study of ionized gas kinematics of single VV galaxies
shows  that their properties drastically depend on their luminosity.
Low luminosity representatives  of  our sample  are in general
metal-deficient objects with $Z$ as low as 1/25--1/10~$Z$\sunn. They
show rather low amplitudes of overall gas motions with superimposed
velocity perturbations, correlated with positions of
enhanced H$\alpha$ emission.
In this aspect they are similar to dwarf irregular and blue compact galaxies.
Luminous VV galaxies appear significantly more evolved  (with $Z$ of
$\sim$1/3--1/2~$Z$\sunn) and with a well traced systematic change
of the line-of-sight velocity with the radial distance, caused by rather
fast rotation.
\item
The extremely low metallicity is now fixed for galaxy VV~499 (DDO~053)
(12+$\log$(O/H) = 7.52$\pm$0.08). This value is
consistent with the estimate obtained
with the empirical method by Skillman et al.~(\cite{Skillman89}).
With its very low metallicity and high $M_{\rm HI}$/$L_{\rm B}$ ratio,
VV~499, along with the dwarf irregular UGC~4483, is probably one of the least
evolved galaxies in the nearby M~81 group.
\item
For all our studied VV galaxies with enhanced star formation, more or less
evident objects exerting sufficiently strong tidal disturbance (that still
in most cases can be classified as a weak tidal) are identified. They include
either sufficiently massive galaxies (as for VV~080, VV~131, VV~523), or a
group at the distance of a few hundred kpc (VV~499). In the case of a very
``isolated'' galaxy (VV~531), there is some evidence of a small sinking
satellite. This indicates that weak interactions can be a trigger mechanism
of SF activity at least in a substantial fraction of VV galaxies studied.
\end{itemize}

\begin{acknowledgements}

The authors appreciate partial financial support from the Russian
federal scientific-technical program (contract 40.022.1.1.1101).
A.Burenkov thanks RFBR for partial support through the grant
01-02-16809. The authors thank the anonymous referee for useful
suggestions, which helped to improve the paper. In
this research the NASA/IPAC Extragalactic Database (NED) was
used, which is operated by the Jet Propulsion Laboratory,
California Institute of Technology, under contract with the
National Aeronautics and Space Administration, and the
Lyon-Meudon Extragalactic Database LEDA.

\end{acknowledgements}



\end{document}